\documentclass[11pt,a4paper]{article}
\usepackage[a4paper, margin=1in]{geometry}
\usepackage{cite}
\usepackage{amsmath,amssymb,amsfonts}
\usepackage{algorithmic}
\usepackage{graphicx}
\usepackage{textcomp}
\usepackage[utf8]{inputenc}
\usepackage{xcolor}
\usepackage{hyperref}
\usepackage{booktabs}
\usepackage{multirow}
\usepackage{bm}
\usepackage{titlesec}

\newcommand{\dotafter}[1]{#1.}
\titleformat{\section}{\normalfont\Large\bfseries}{\thesection}{1em}{\dotafter}
\titleformat{\subsection}{\normalfont\large\bfseries}{\thesubsection}{1em}{\dotafter}
\titleformat{\paragraph}[runin]{\normalfont\normalsize\bfseries}{\theparagraph}{0pt}{\dotafter}

\graphicspath{{./images/}}

\newcommand{\xx}{\bm{x}}
\newcommand{\yy}{\bm{y}}

\newcommand{\angles}{\bm{\psi}}
\newcommand{\Angles}{\Psi}
\newcommand{\angl}{\psi}
\newcommand{\A}{\mathcal{A}}

\begin{document}

\title{Conditional Diffusion Posterior Alignment for Sparse-View CT Reconstruction}

\author{
  Luis Barba\thanks{Swiss Data Science Center (SDSC) in Paul Scherrer Institute (PSI), Villigen, Switzerland. E-mail: \texttt{luis.barba-flores@psi.ch}.} \and
  Johannes Kirschner\thanks{Swiss Data Science Center (SDSC) and ETH Zurich, Switzerland. E-mail: \texttt{johannes.kirschner@ethz.ch}.} \and
  Benjam\'in B\'ejar\thanks{Swiss Data Science Center (SDSC) and Paul Scherrer Institute (PSI), Villigen, Switzerland. E-mail: \texttt{benjamin.bejar@psi.ch}.}
}

\date{}

\maketitle

\begin{abstract}
Computed Tomography (CT) is a widely used imaging modality in medical and industrial applications. To limit radiation exposure and measurement time, there is a growing interest in sparse-view CT, where the number of projection views is significantly reduced. Deep neural networks have shown great promise in improving reconstruction quality in sparse-view CT, especially generative diffusion models. However, these methods struggle to scale to large 3D volumes due to several reasons: (i) the high memory and computational requirements of 3D models, (ii) the lack of large 3D training datasets, and (iii) the inconsistencies across slices when using 2D models independently on each slice.
We overcome these limitations and scale diffusion-based sparse-view CT reconstruction to large 3D volumes by combining conditional diffusion with explicit data consistency. We propose Conditional Diffusion Posterior Alignment (CDPA) to enable scalable 3D sparse-view CT reconstruction. A 2D U-Net diffusion model is conditioned on an initial 3D reconstruction to improve inter-slice consistency, combined with data-consistency alignment to match measured projections. 
Experiments on synthetic and real Cone Beam CT (CBCT) data show state-of-the-art performance, with ablations that confirm the synergistic effects of the proposed pipeline.
Finally, we show that the same principles also strengthen fast denoising U-Nets, yielding near-diffusion quality at a fraction of the computational cost.
\end{abstract}


\noindent\textbf{Funding:} This project acknowledges funding support from the SDSC CHIP project under grant no.~C22-11L.


\section{Introduction}\label{sec:introduction}

Computed Tomography (CT) is a non-invasive imaging technique that reconstructs cross-sectional images of an object from its X-ray projections.
It has become an essential tool in medical diagnostics, industrial inspection, and scientific research.
However, acquiring high-quality CT images often requires a large number of projection views, leading to increased radiation dose and, in some settings, to longer scan times. \emph{Sparse-view} CT aims to reduce the number of projections acquired while maintaining image quality, but it remains a challenging inverse problem due to the ill-posed nature of the reconstruction task.

When the number of projection angles is drastically reduced, to lower radiation dose or scan time, classical analytical algorithms such as \emph{filtered back-projection} (FBP) and its Cone Beam CT (CBCT) counterpart, the \emph{Feldkamp-Davis-Kress} (FDK) algorithm, produce severe streak artifacts. In the particular case of CBCT where we focus our experimental results, the reconstructions cannot be decomposed into independent slices as in parallel-beam geometry CT. Therefore, CBCT poses an intrinsically 3D problem, where the aforementioned artifacts span across different slices, and where 2D approaches tend to be insufficient.
Early iterative reconstruction schemes with hand-made regularizers (e.g.,\ total-variation) improve these artifacts but require long run-times, are sensitive to hyper-parameter tuning, and still fall short under high-sparsity scenarios.

Deep learning has transformed sparse-view CT by introducing data-driven priors that learn image statistics directly from examples. Supervised convolutional neural networks (CNNs), in particular U-Net-based models such as FBPConvNet \cite{Jin2017}, have emerged as the standard reference method when used as post-processing on FBP/FDK reconstructions, providing good artifact suppression. 
We refer to these models as \emph{FBP/FDK-denoisers}. More recent approaches have improved reconstruction quality through geometry-aware encoders \cite{liu_geometry-aware_2025} or sliding 3D sub-volumes \cite{zhang_sliding_2025}, setting new state-of-the-art benchmarks for CBCT. In parallel, unsupervised methods such as Deep Image Prior \cite{Ulyanov2018} and score-based diffusion models \cite{chung2022diffusion,song2023solving,barba2025diffusion} have emerged as powerful priors achieving state-of-the-art results. However, these methods are computationally expensive, making it prohibitive to scale them to large 3D-volumes. This leaves us with slice-by-slice approaches using 2D models that out-of-the-box suffer from inconsistencies across slices in 3D reconstructions.

Two strategies can restore 3D consistency to such a slice-by-slice 2D diffusion prior. The first imposes an explicit, hand-crafted regularizer coupling adjacent slices directly inside the sampling process, e.g.\ a total-variation penalty solved via ADMM, as in DiffusionMBIR~\cite{chung2023solving3d}. The second, which we adopt, inherits 3D-consistent behavior by conditioning the diffusion model on an initial 3D reconstruction. We show in Sec.~\ref{sec:mbir_comparison} that although the two strategies reach comparable PSNR and SSIM, the explicit-regularization route is not a free alternative: it over-smooths the axial axis relative to the anatomy's own structure, reduces the sampler's posterior variance, undermining the benefit of posterior averaging via the $\mu(\cdot)$ operator, and produces substantially worse perceptual quality. Our conditioning strategy matches its fidelity without incurring any of these three costs.

\subsection{Contributions}
Motivated by these limitations, we propose the use of conditioning on slices of the FBP/FDK volume reconstructions to improve 3D consistency. We present two complementary 2D U-Net-based approaches for sparse-view CT: (i) Conditional Diffusion models with Posterior Alignment (CDPA) and (ii)~refined FBP/FDK-denoising networks, both with slice-positional encoding and data-consistency alignment.
Although our method applies to all CT geometry settings, we evaluate them in CBCT datasets that deal with an intrinsically 3D reconstruction problem, where sparse-view artifacts span different slices, and which is a more prominent setting in the medical community.
Our key contributions are as follows.
\begin{itemize}
    \item We demonstrate that conditioning on FBP/FDK reconstructions significantly improves the quality and 3D slice consistency of diffusion model output, addressing a known limitation of unconditional slice-by-slice approaches.
    \item We introduce CDPA, a conditional ReSample-style diffusion posterior-alignment framework for large-volume sparse-view CT. We show that it systematically outperforms unconditional diffusion on all benchmarked baselines.
    \item We show that a simple Monte Carlo averaging strategy over diffusion posterior samples yields substantial PSNR and, surprisingly, also SSIM improvements, suggesting near-unimodal posteriors in practice.
    \item We show that an alternative, explicit-regularization strategy for enforcing 3D consistency (DiffusionMBIR's ADMM+TV guidance, applied to our own pretrained checkpoints with no retraining) reaches comparable PSNR/SSIM to CDPA, but at the cost of axial over-smoothing, a collapsed posterior, and substantially worse perceptual quality -- costs that our conditioning avoids entirely.
    \item We revisit the FBPConvNet paradigm \cite{Jin2017} with modern U-Net architectures incorporating cross-attention and slice-position conditioning for better artifact removal. Combined with inference-time data-consistency fine-tuning, this approach nearly matches diffusion performance at a fraction of the computational cost and surpasses all previous approaches.
    \item We provide extensive experiments across synthetic and real-world CBCT datasets, demonstrating state-of-the-art PSNR and SSIM compared to recent methods.
\end{itemize}

The code supporting the results in this paper can be found in~\url{https://github.com/SwissDataScienceCenter/cbct_cdpa}.

\section{Related Work}\label{sec:related_work}

\paragraph{Classical and iterative reconstruction}
The filtered back-projection (FBP) algorithm is the classical analytical solution in parallel-beam CT, exploiting the Fourier slice theorem for exact inversion under dense angular sampling. Its cone-beam counterpart, the Feldkamp-Davis-Kress (FDK) algorithm, serves the same role in CBCT. Under aggressive sparse-view or low-dose regimes, however, both FBP and FDK violate their sampling assumptions and produce globally oriented coherent streak artifacts that obscure fine structure.
Algebraic reconstruction techniques (ART) and variational formulations (e.g.\ quadratic data fidelity plus sparsity or total-variation penalties) alleviate streaking by explicitly regularizing the ill-posed inverse. Yet they entail (i)~many forward/backprojection sweeps, (ii)~sensitive balancing between data fidelity and regularization strength, and (iii)~difficulty adapting the prior to heterogeneous textures (e.g.\ simultaneously preserving sharp edges and smooth backgrounds). As sparsity increases, hand-crafted priors either over-smooth or leave residual structured artifacts, motivating learned data-driven priors.

\paragraph{Supervised CNN post-processing}
A widely adopted pattern is \emph{FBP+CNN}: compute a fast baseline reconstruction (FBP or FDK) and feed it to a convolutional network that maps streak-corrupted images to clean images. Jin \emph{et\,al.}\ \cite{Jin2017} demonstrated that a moderately sized U-Net (FBPConvNet) can remove severe streaks from extremely sparse sinograms (e.g.,\ tens of views) in near real-time per slice while outperforming total-variation iterative solvers. We refer to 
models following this paradigm as \emph{FBP/FDK-denoisers}. Subsequent variants refined network depth, multi-scale skip connections, loss functions, and training curricula, cementing U-Nets as a strong supervised baseline. 

Recent state-of-the-art methods for CBCT have incorporated geometric awareness and 3D processing. GAAL \cite{liu_geometry-aware_2025} uses a geometry-aware encoder-decoder framework that encodes multi-view 2D features from various 2D X-ray projections with a 2D CNN encoder, back-projects the multi-view 2D features into 3D space to formulate a comprehensive volumetric feature map, followed by a 3D CNN decoder to recover the 3D CBCT image. S-STAR Net \cite{zhang_sliding_2025} tackles sparse-view reconstruction with sliding 3D sub-volumes, a difference-enhancement loss, volume-attention residuals, and Fourier-transform convolutions, achieving state-of-the-art PSNR/SSIM on walnut and CQ500 datasets under 30-view acquisition.
Other supervised 3D baselines include PTNet3D~\cite{zhang2022ptnet3d}, which learns a direct 3D mapping for sparse-view reconstruction, and Pix2Pix3D~\cite{isola2017image}, which adapts conditional image-to-image translation (pix2pix) to 3D volumes.
All these 3D-algorithms are, however, limited to low-resolution volumes due to memory constraints, and do not scale to the high-resolution CBCT volumes that we target in this work.

\paragraph{Unsupervised and zero-shot methods}
In scenarios where large paired sinogram--image datasets are unavailable, \emph{zero-shot} methods exploit architectural or generative priors without external training data. Deep Image Prior (DIP) \cite{Ulyanov2018} reconstructs a single scan by optimizing the weights of a randomly initialized CNN to fit the measurements, exploiting architectural bias alone as an implicit prior. The network, fed with random noise as input, is iteratively updated to minimize data fidelity; the resulting image often converges to a plausible solution before overfitting measurement noise. Although effective, DIP incurs substantial per-scan optimization overhead. Warm-start strategies \cite{Barbano2023} pre-train the network on synthetic samples and then fine-tune per instance, reducing iteration counts and stabilizing optimization. An ``educated warm start'' approach uses FBP reconstructions as network input instead of pure noise, further accelerating convergence. Still, DIP-style methods remain computationally heavy and do not scale well to large 3D volumes, making them less suited for medical settings and high-resolution scientific imaging.

\paragraph{Diffusion models for inverse problems}
Score-based diffusion models \cite{ho2020denoising} have emerged as powerful generative priors, learning time-indexed score functions of noisy data distributions and enabling high-fidelity sampling decoupled from specific forward operators. For inverse problems, a general strategy iteratively denoises from pure noise while nudging samples toward measurement consistency. 

Diffusion Posterior Sampling (DPS) \cite{chung2022diffusion} formalized this conditioning as adding an approximate likelihood gradient to the learned score, improving stability under noise and allowing flexible noise models without retraining the diffusion prior. DDRM (Denoising Diffusion Restoration Models) \cite{Kawar2022} leverages a spectral factorization to incorporate measurement information analytically for certain linear operators, though computing or storing such decompositions is impractical for large CT system matrices.

Latent diffusion with hard data consistency (e.g.,\ ReSample \cite{song2023solving}) extends these ideas to work with latent image representations. It introduces an optimization-based latent refinement that enforces exact forward consistency before a resampling step restores adherence to the learned latent manifold. 
Recent work \cite{barba2025diffusion} has adapted these ideas to non-latent diffusion for CT, further improving reconstruction quality. Deterministic samplers such as DDIM (Denoising Diffusion Implicit Models) \cite{DDIM2020} reduce wall-clock time by using non-Markovian trajectories with fewer function evaluations, which is critical given the computational burden of repeated forward/backprojection inside each reverse diffusion step.

Closely related to our work, DOLCE \cite{liu2023dolce} introduced conditional diffusion models for limited-angle CT reconstruction, integrating data-consistency updates with the sampling steps of a diffusion model conditioned on the limited-angle measurements. DOLCE enforces measurement consistency via proximal gradient steps at each reverse diffusion iteration. Our CDPA method is conceptually related, but differs in targeting large-volume sparse-view CBCT and in using a ReSample-style posterior-alignment approach \cite{song2023solving} rather than proximal gradients, combined with FDK-conditioning and slice-position conditioning for 3D consistency.

Despite their success, when applied slice-by-slice with 2D models to 3D volumes, diffusion methods can produce inter-slice inconsistencies because each slice is processed independently without enforcing 3D coherence. While 3D diffusion models could address these limitations in theory, in practice their memory and computational demands are prohibitive at high resolutions, not to mention the increased amount of training data required.
Two lines of work address this with 2D priors only: Lee et al.~\cite{lee2023improving} attempted to address this issue by modeling the 3D data distribution as a product of 2D distributions sliced in different axes and using them alternately during reconstruction, and DiffusionMBIR~\cite{chung2023solving3d} augments a 2D diffusion prior with a one-dimensional total-variation penalty along the remaining axis inside an ADMM data-consistency step.
We compare directly against DiffusionMBIR's guidance mechanism in Sec.~\ref{sec:mbir_comparison}, applying it to our own pretrained DPA/CDPA checkpoints with no retraining. Despite reaching comparable but lower PSNR/SSIM to CDPA there, we find it introduces three concrete costs -- axial over-smoothing, a collapsed posterior, and substantially worse perceptual quality -- that FDK-conditioning avoids.

\paragraph{Summary and positioning}
Supervised CNNs exhibit good performance but fall short of state-of-the-art performance. Zero-shot methods (DIP) remove dependency on large datasets but incur high per-instance cost and do not exploit data priors, thus being surpassed by models using learned priors. Diffusion models unify strong generative fidelity with flexible conditioning, and recent advances (DPS, latent consistency, accelerated samplers) make them increasingly practical for CT. However, scaling diffusion methods to large 3D volumes remains challenging due to memory, compute, and training data limitations, leading to slice-by-slice approaches that suffer from inter-slice inconsistencies.

Our work centers on exploiting diffusion posterior structure and data-consistency refinement under sparse-view regimes. We introduce Conditional Diffusion Posterior Alignment to scale 2D-diffusion approaches to 3D volumes using data-consistency alignment, as well as slice-position and FDK/FBP conditioning. We also show that these techniques can be applied to FBP/FDK-denoisers elevating them to near-diffusion performance without major architectural changes.
Our 2D-based methods scale to much higher resolution than intrinsic 3D models, due to the independent processing of slices.

\section{Method} \label{sec:Method}
In this section, we describe the methods used in our experiments. We focus on two main approaches: diffusion-based models and FBP/FDK-denoising models, both based on the U-Net architecture.

\paragraph{Problem setup and notation}
We denote by $\xx_0\in\mathbb{R}^N$ the (unknown) target image and by $\A:\mathbb{R}^N\to\mathbb{R}^M$ the (discrete) forward operator (parallel beam Radon transform or Cone Beam Projection). We index projection views by an angle $\angl\in\Angles$, where the finite acquired set is $\angles=\{\angl_1,\ldots,\angl_{K}\}$. We decompose $\A$ into per-angle operators $\A_{\angl}$ so that the full set of projections is
\[
\yy=\{\yy_{\angl}\}_{\angl\in\angles},\qquad \yy_{\angl}=\A_{\angl}\xx_0+\epsilon_{\angl},
\]
with $\epsilon_{\angl}$ measurement noise (assumed i.i.d.\ Gaussian). Thus $\yy_{\angl}$ denotes the detector readings for view $\angl$.

\subsection{Score-based diffusion priors and posterior sampling}
The main idea of our method is to train a diffusion model to learn the target image distribution, and then use data-consistency guidance to align the diffusion sample with the measurements as done in~\cite{barba2025diffusion}. However, in addition to this guidance step, we make the diffusion model conditional by passing the sparse FDK/FBP reconstruction as input to approximate the conditional score function. Our approach combines these two conditional sampling approaches for superior performance. We describe this method in detail below.

We adopt score-based diffusion models as learned image priors. Given a clean sample $\xx_0$, we construct a noisy $\xx_t$ at noise level (or time) $t\in[0,1]$ via $\xx_t = \alpha_t \xx_0 + \sigma_t \epsilon$, $\epsilon\sim\mathcal{N}(0,I)$ with known scalar schedules $(\alpha_t,\sigma_t)$ (e.g., derived from a variance schedule in DDPM (Denoising Diffusion Probabilistic Models) \cite{ho2020denoising}).
The marginal $p_t(\xx_t)$ denotes the probability density of the random variable $\xx_t$ at time $t$ when $\xx_0$ is drawn from the empirical data distribution. A \emph{score network} $s_\theta(\xx_t,t)$ is trained to approximate $\nabla_{\xx_t} \log p_t(\xx_t)$ (the \emph{score} of the noisy marginal) by minimizing a denoising-score matching objective. We use the common $\epsilon$-parameterization that predicts the added noise $\epsilon_\theta(\xx_t,t)$; the two are related by a scaling factor. From a network output we form a denoised (Tweedie) estimate $\hat{\xx}_0(\xx_t)$ of $\xx_0$ which we will later use inside a data-consistency loss function. When conditioning on measurements $\yy$ our goal is to sample from the posterior distribution $p(\xx_0\mid \yy)$.
At inference, we use a fast sampler based on DDIM \cite{DDIM2020}.

\paragraph{Diffusion posterior alignment (DPA)} For reconstruction we target the posterior over clean images given measurements $p(\xx_0\mid \yy)$. Using the diffusion prior, we can express the posterior at time $t$ as $p_t(\xx_t\mid \yy)$ and its score as:
\begin{equation}
\nabla_{\xx_t}\log p_t(\xx_t\mid \yy)
=
\nabla_{\xx_t}\log p_t(\xx_t) + \nabla_{\xx_t}\log p_t(\yy\mid \xx_t).
\label{eq:dc_score}
\end{equation}
Diffusion Posterior Sampling (DPS) \cite{chung2022diffusion} approximates the term $\nabla_{\xx_t}\log p_t(\yy\mid \xx_t)$ by propagating a data-consistency gradient from a current denoised estimate $\hat{\xx}_0(\xx_t)$ (e.g.,\ via Tweedie's formula) back to $\xx_t$. We define the per-step data-consistency loss
\begin{equation}
L_{\text{DC}}(\xx_t)
=
\sum_{\angl\in\angles}\big\|\A_{\angl}\hat{\xx}_0(\xx_t)-\yy_{\angl}\big\|_2^2.
\label{eq:dc-loss}
\end{equation}
DPS~\cite{chung2022diffusion} uses $\nabla_{\xx_t} L_{\text{DC}}$ (through $\hat{\xx}_0$) as a likelihood score surrogate (scaled by an empirically tuned guidance weight). Each reverse diffusion update is then nudged toward the measurement manifold while the prior score preserves realism. This however requires to backpropagate through the diffusion model, which is computationally expensive, and is not compatible with the DDIM sampler we want to use.
Instead, we use a variant introduced by \cite{song2023solving} that takes the gradient $\nabla_{\hat{\xx}_0(\xx_t)} L_{\text{DC}}$ with respect to the Tweedie's formula estimation $\hat{\xx}_0(\xx_t)$. That is, we stay in image space without backpropagating the gradient to the noisy manifold on $\xx_t$. After performing several gradient steps using $\nabla_{\hat{\xx}_0(\xx_t)} L_{\text{DC}}$, ensuring data consistency, we map back the resulting image to the noisy manifold at time $t$. To this end, we use the ReSample algorithm introduced by \cite{song2023solving}, but avoiding the use of latent space diffusion models. Our approach is the same as in \cite{barba2025diffusion}, but we extend it to CBCT geometries in this paper. We refer to this unconditional diffusion posterior sampling with data consistency alignment method as \emph{Diffusion Posterior Alignment} (DPA).

\paragraph{Conditional diffusion posterior alignment (CDPA)}
In the conditional formulation, we use our neural network to approximate directly the conditional score function by training it to predict the score of the conditional distribution with $\yy$ as input in the form of its FBP/FDK reconstruction, i.e.,
\begin{equation}
\nabla_{\xx_t}\log p_t(\xx_t\mid \yy) \approx \nabla_{\xx_t}\log p_t(\xx_t\mid \mathrm{FDK}(\yy)).
\label{eq:score_cond}
\end{equation}

We end thus with two ways to approximate the posterior score: the conditional score network and the data-consistency gradient. Our final posterior score approximation is given by the combination of both~(\ref{eq:dc_score}) and (\ref{eq:score_cond}). We achieve this by using the conditional diffusion model to obtain the estimate $\hat{\xx}_0$ through Tweedie's formula, and then applying data-consistency gradient descent updates to enforce data consistency on this estimate before mapping back to the noisy manifold at time $t$ using the aforementioned ReSample step.
Throughout this work, we refer to this method as \emph{Conditional Diffusion Posterior Alignment} (CDPA).

Technically, one could use only the conditional score network without the data-consistency alignment step. However, we found that not using the data-consistency alignment step results in much inferior reconstruction quality and reconstructions that are almost never data-consistent.

\paragraph{Slice conditioning}
We noticed that due to the cone-beam geometry of CBCT, artifacts generated by FDK in the sparse-view setting differ across regions of the reconstructed volume. This led us to design a model that is position-aware and is capable of handling artifacts differently in different parts of the volume.
To this end, we add an additional conditioning to the U-Net behind the diffusion model by using the slice-index of the reconstructed volume.
More specifically, we used the cross-attention layers of the U-Net architecture to attend to a positional encoding of this slice index in the bottleneck layers of the U-Net model.

\subsection{FBP/FDK-denoising models}
We revisit the ideas of \cite{ronneberger2015u,Jin2017} but using larger modern U-Net architectures with cross-attention layers. The network takes as input an initial FBP/FDK reconstruction $\hat{\xx}_{\text{FBP}}$ obtained from the projections $\yy$ and produces as output a denoised image $\hat{\xx}_0$.
The network is trained to minimize the difference between $\hat{\xx}_0$ and the corresponding ground-truth image $\xx_0$ in the $\ell_2$ sense. The training dataset consists of pairs of sinograms and ground-truth images, coming from the same distributions as the images used in the testing phase.
In addition to the FBP/FDK reconstruction, we also condition the network on the slice index within the volume, following the same positional encoding strategy as in transformer models. This spatial-aware conditioning helps the network to better adapt to the varying artifact patterns that depend on the slice position which are common in CBCT reconstructions.

We notice that $\hat{\xx}_0$ is almost never data consistent. That is, the output $\hat{\xx}_0$ does not satisfy the data-consistency condition $\A\hat{\xx}_0=\yy$. This is a common issue with FBP/FDK-denoising models, as they focus on denoising the FBP reconstruction without explicitly enforcing consistency with the measured data.
Rather than adding an additional data-consistency term to the loss function during training, which did not produce satisfactory results, we propose a simple fine-tuning step at inference time to enforce data-consistency.
We project $\hat{\xx}_0$ onto the data-consistency manifold by minimizing the data-consistency loss $L_{\text{DC}}$ from equation~(\ref{eq:dc-loss}) for a few iterations using a small learning rate with the Adam optimizer. This step refines the reconstruction by enforcing consistency with the measured data $\yy$.

\section{Datasets}
\label{sec:datasets}

To evaluate the effectiveness of our methods, we conducted CBCT experiments using both simulated datasets (dental and spine~\cite{deng2021ctspine1k}) and a real-world dataset (walnut~\cite{der2019cone}).
Our low-resolution experiments ($256^3$ voxel volumes) follow the same setup and evaluation metrics as in~\cite{liu_geometry-aware_2025} for fair comparison with recent state-of-the-art methods. These low-resolution datasets are available on Hugging Face at \url{https://huggingface.co/datasets/Zhentao-Liu/TMI2024_SVCT_dataset}.
Our high-resolution experiments on the full resolution walnut dataset~\cite{der2019cone} ($501^3$ voxel volumes) further demonstrate the scalability and effectiveness of our methods in more challenging scenarios. We summarize the datasets, preprocessing steps, and evaluation metrics below.

\subsection{CBCT Datasets}

\paragraph{Walnut (real CBCT)}
We adopt 42 real walnut scans, split 37/5~\cite{der2019cone}. This dataset consists of three scans at different heights (high/middle/low), each with 1200 views at $0.3^\circ$ steps, with recorded flat-field ($I_1$) and dark-field ($I_0$). A reference \emph{ground-truth} CBCT is reconstructed by combining the three height scans using an iterative procedure.
We work with two resolution settings for this dataset. The first, at low resolution, allows us to compare our method with baselines that struggle to scale to full resolution.
Following exactly~\cite{liu_geometry-aware_2025}, raw projections are downsampled from $768{\times}972$ with an effective detector pixel size of $0.1496~\mathrm{mm}$ to a third of the size, resulting in projections of size $256{\times}324$ and reference volumes with $256^3$ voxels. For model inputs we use uniformly spaced $N=20$ views from the \emph{middle} scan.

We also work with full resolution projections resulting in reference volumes with $501^3$ voxels. In this setting, we restrict our models to work with the more challenging \emph{high} scan (1200 views at $0.3^\circ$ steps). This produces more artifacts to further highlight the efficiency of our methods. Inputs use $N\in\{20, 40, 60, \ldots, 160, 180\}$ uniformly spaced views to test the effect of varying sparsity.
To compare directly with S-STAR Net~\cite{zhang_sliding_2025}, we also train and test with $N =30$ views. However, since their algorithm works only with a crop due to computational resources, we crop our $501^3$ reconstructed volumes to a central crop with resolution $360 \times 360 \times 150$ to exactly match their setting.

\paragraph{Dental (simulated)}
For this dataset, we follow the exact dataset from~\cite{liu_geometry-aware_2025}.
They used 130 dental CBCT volumes from patients resampled to $256^3$ voxels at $(0.3133~\mathrm{mm})^3$, split 100/10/20 for train/val / test. For each case, we generate synthetic X-ray projections with a circular cone-beam trajectory using the ASTRA toolbox\cite{van2015astra,van2016fast,palenstijn2011performance}: $N=20$ uniformly spaced views over $[0^\circ,360^\circ)$, detector size $256{\times}256$ with pixel spacing $0.4386~\mathrm{mm}$, source--object distance $L_{sb}=500~\mathrm{mm}$, and object--detector distance $L_{bd}=200~\mathrm{mm}$~\cite{liu_geometry-aware_2025}.
Both the spine and dental datasets are third-party, de-identified public releases. CTSpine1K states that all volumes are de-identified in accordance with the IRB policies of its contributing sites~\cite{deng2021ctspine1k}; the dental CBCT volumes are redistributed under the GAAL/TMI2024 release~\cite{liu_geometry-aware_2025} under its original license. The present study performed no new human-subject data acquisition and required no additional ethical approval.

\paragraph{Spine (CTSpine1K-derived, simulated)}
For this dataset, we follow the exact dataset from~\cite{liu_geometry-aware_2025}.
They used 130 CT volumes from CTSpine1K~\cite{deng2021ctspine1k}, resample/crop/pad each to $256^3$ with $(2~\mathrm{mm})^3$ voxels, split 100/10/20, and mirror the dental view protocol ($N=20$ uniform views) to generate synthetic projections. These projections use a $256{\times}256$ detector with $3~\mathrm{mm}$ pixels, $L_{sb}=1000~\mathrm{mm}$ and $L_{bd}=500~\mathrm{mm}$. This protocol is noiseless, while the walnut experiments use real, noisy measurements.

\paragraph{Projection correction for real data (walnut)}
Raw detector counts $I$ are converted to line integrals via Beer--Lambert with per-pixel flat/dark correction:
\[
P = -\ln\!\left(\frac{I-I_0}{I_1-I_0}\right),
\]
where $I_0$ and $I_1$ are the dark-field and flat-field measurements, respectively.

\begin{table}[tbp]
\centering
\caption{Quantitative comparison of reconstruction methods across three CBCT datasets ($n=20$ test volumes for Dental and Spine, $n=5$ for Walnut). Values are reported as mean ± standard deviation.
DPA stands for Diffusion Posterior Alignment, CDPA for Conditional Diffusion Posterior Alignment, and FT for Fine-Tuning. The operator $\mu$ indicates the mean of 10 samples from the posterior distribution. Runs with $^*$ indicate results taken from GAAL's paper~\cite{liu_geometry-aware_2025}; DIF-Net and GAAL (unstarred) were re-run by us using the authors' released checkpoints/code under our evaluation protocol, all other unstarred rows are our own methods.
The best value per column is in \textbf{bold}.}
\label{tab:results_comparison}
\resizebox{\textwidth}{!}{%
\begin{tabular}{l|cc|cc|cc}
\toprule
\multirow{2}{*}{\textbf{Method}} & \multicolumn{2}{c|}{\textbf{Dental (20 views)}} & \multicolumn{2}{c|}{\textbf{Spine (20 views)}} & \multicolumn{2}{c}{\textbf{Walnut (20 views)}} \\
\cmidrule(lr){2-3} \cmidrule(lr){4-5} \cmidrule(lr){6-7}
& \textbf{PSNR} & \textbf{SSIM} & \textbf{PSNR} & \textbf{SSIM} & \textbf{PSNR} & \textbf{SSIM} \\
\midrule
FDK & 22.54±0.57 & 0.430±0.021 & 22.77±1.43 & 0.381±0.040 & 16.07±0.22 & 0.194±0.004 \\

GD$_{zero}$ & 27.68±0.95 & 0.785±0.015 & 27.77±2.18 & 0.868±0.039 & 23.93±0.39 & 0.590±0.008 \\
GD$_{FDK}$ & 27.97±0.46 & 0.745±0.018 & 28.34±1.55 & 0.821±0.051 & 23.36±0.24 & 0.582±0.006 \\

NAF~\cite{zha2022naf}$^*$ & 28.77±0.86 & 0.793±0.020 & 30.80±1.93 & 0.912±0.024 & 25.36±0.22 & 0.680±0.006 \\
SCOPE3D~\cite{gu2024fpga}$^*$ & 29.39±0.87 & 0.807±0.019 & 30.91±2.00 & 0.908±0.024 & 25.92±0.19 & 0.714±0.009 \\
SNAF~\cite{fang2022snaf}$^*$ & 30.93±0.51 & 0.844±0.015 & 31.69±1.39 & 0.902±0.035 & 27.43±0.29 & 0.748±0.006 \\
PatRecon~\cite{shen2019patient}$^*$ & 19.95±0.85 & 0.569±0.038 & 18.57±1.75 & 0.641±0.047 & 17.30±0.55 & 0.699±0.033 \\
PixelNeRF~\cite{yu2021pixelnerf}$^*$ & 26.85±0.57 & 0.775±0.014 & 28.68±1.64 & 0.890±0.033 & 24.79±0.15 & 0.732±0.004 \\
DIF-Net~\cite{lin2023learning} & 30.48±0.80 & 0.870±0.015 & 32.28±1.75 & 0.901±0.034 & 25.99±0.14 & 0.707±0.004 \\
GAAL~\cite{liu_geometry-aware_2025} & 31.44±1.00 & 0.891±0.016 & 33.14±1.88 & 0.938±0.020 & 26.84±0.37 & \textbf{0.889±0.006} \\

\midrule
DPA & 30.82±0.80 & 0.831±0.021 & 32.85±1.66 & 0.892±0.028 & 25.34±0.26 & 0.648±0.011 \\
CDPA & 33.32±0.69 & 0.890±0.013 & 35.84±1.74 & 0.939±0.020 & 29.58±0.34 & 0.795±0.012 \\
FDK-denoiser & 33.07±0.90 & 0.901±0.013 & 36.62±1.69 & 0.950±0.018 & 28.18±0.98 & 0.888±0.005 \\
\midrule
$\mu$(DPA) & 33.61±0.77 & 0.906±0.012 & 35.11±1.68 & 0.942±0.014 & 26.21±0.27 & 0.666±0.006 \\
$\mu$(CDPA) & \textbf{34.72±0.70} & \textbf{0.921±0.010} & \textbf{37.08±1.83} & \textbf{0.952±0.016} & \textbf{30.50±0.83} & 0.813±0.014 \\
FDK-denoiser + FT & 33.30±0.72 & 0.882±0.012 & 36.90±1.63 & 0.944±0.024 & 28.90±0.54 & 0.740±0.014 \\
\bottomrule
\end{tabular}%
}
\end{table}

\section{Experiments}
\label{sec:experiments}

For each of the datasets described in Section~\ref{sec:datasets}, we train both 2D diffusion-based models and FBP/FDK-denoising models using the methods described above and the corresponding training splits.
The low-resolution datasets are restricted to $N=20$ views, while for the high-resolution walnut dataset we experiment with varying numbers of views $N\in\{20,40,60,\ldots,160,180\}$. For this high-resolution setting, we train a single model for all view counts by randomly selecting the number of views at each training iteration. This yields a single model that can reconstruct images from different levels of sparsity without training a separate model for each case. Our model then includes an additional conditioning input on the number of views, allowing it to differentiate between sparsity levels.
For the specific case of $N=30$ views, we fine-tuned the model with samples at this specific sparsity to enable a fair comparison with S-STAR Net~\cite{zhang_sliding_2025}. We also use this setting to compare our method with explicit-regularization (DiffusionMBIR); see below.

To train our diffusion model, we use the standard MSE loss on the noise prediction as in \cite{ho2020denoising} for both the unconditional and conditional models. For the FBP/FDK-denoising models, we use the $\ell_2$ loss between the denoised output and the ground-truth image. We use the Adam optimizer for all our experiments. More details about the training hyperparameters and architectures are provided in the supplementary material~\ref{sec:supplementary}.

\subsection{Baseline Methods}

We compare our proposed methods against several state-of-the-art baselines for sparse-view CT reconstruction. For the $256^3$ resolution CBCT experiments, we compare against all baselines reported in GAAL~\cite{liu_geometry-aware_2025}.

\paragraph{FDK} The Feldkamp-Davis-Kress (FDK) algorithm serves as the analytical baseline for CBCT reconstruction. Under extreme sparse-view conditions, FDK produces severe streak artifacts that motivate data-driven approaches.

\paragraph{Iterative methods} We include gradient descent-based iterative reconstruction (GD) using data-consistency terms but lacking learned priors. We benchmark two variants of this approach, one initialized with zeros, GD$_{zero}$, and a second initialized with the FDK reconstruction, GD$_{FDK}$.
Gradients are computed with respect to~(\ref{eq:dc-loss}). To this end, we developed a PyTorch wrapper in the ASTRA toolbox~\cite{van2015astra,van2016fast,palenstijn2011performance} that allows us to compute projections and backpropagate gradients to image space. We use the Adam Stochastic Gradient Descent optimizer with mini-batches of 30 projections to improve performance and speed. Our fine-tuning steps and alignment diffusion guidance are implemented by calling this GD module.

\paragraph{Supervised CNN baselines} We compare against several recent approaches reported in GAAL \cite{liu_geometry-aware_2025}: PatRecon~\cite{shen2019patient} and SCOPE3D (a 3D convolutional approach)~\cite{gu2024fpga}.

\paragraph{Neural rendering methods} We include NAF (neural attenuation fields)~\cite{zha2022naf}, SNAF (sparse-view neural attenuation fields)~\cite{fang2022snaf} and PixelNeRF~\cite{yu2021pixelnerf} which leverage neural rendering techniques for 3D reconstruction.
The results of these methods are taken from \cite{liu_geometry-aware_2025}.

\paragraph{Recent state-of-the-art} We compare against two recent state-of-the-art methods: GAAL \cite{liu_geometry-aware_2025}, which employs a geometry-aware encoder-decoder framework with multi-view 2D features and 3D volumetric reconstruction, and DIF-Net~\cite{lin2023learning}, a deep image fusion network for sparse-view CT.

\paragraph{Explicit-regularization guidance (DiffusionMBIR)} We also compare against DiffusionMBIR~\cite{chung2023solving3d}, which augments a \emph{pretrained, unconditional} 2D diffusion prior with an ADMM data-consistency step that enforces an explicit 1D total-variation penalty coupling adjacent slices along the axial axis, solved via alternating CG least-squares updates and soft-thresholding at every reverse-diffusion timestep. We implement this guidance mechanism as a drop-in replacement for our own alignment guidance (Sec.~\ref{sec:Method}), which lets us apply it directly to our existing pretrained DPA and CDPA checkpoints without retraining: DPA+ADMM-TV reproduces DiffusionMBIR, while CDPA+ADMM-TV additionally retains our FDK-conditioning, letting us test whether the two consistency mechanisms are complementary. Because the ADMM penalty weight and iteration budget require per-dataset tuning, we restrict this comparison to the walnut dataset at $N=30$ views, full $501^3$ resolution -- the real-measurement, high-resolution regime central to this paper's scaling claim (Sec.~\ref{sec:mbir_comparison}).

All baseline results marked with $^*$ in Table~\ref{tab:results_comparison} are taken directly from the GAAL paper \cite{liu_geometry-aware_2025} to ensure fair comparison under identical evaluation protocols.

For the $501^3$ resolution CBCT experiments, we compare our methods against FDK and gradient descent-based iterative reconstruction (GD) baselines only, as most other methods struggle to scale to this higher resolution.
For the case of 30 views, we compare our results with S-STAR Net~\cite{zhang_sliding_2025} and the baselines presented there, which is the only method, together with DiffusionMBIR, that we are aware of that scales close to this resolution, though S-STAR Net only operates on a center crop of the full-resolution walnut volume.
Since no code was provided for S-STAR Net, we compare against the results reported in their paper. Besides S-STAR-Net, we include their results on PTNet3D~\cite{zhang2022ptnet3d} and Pix2Pix3D~\cite{isola2017image} in our comparison.

\subsection{Evaluation Metrics}\label{sec:metrics}

We evaluate the reconstruction quality in sparse-view tomography using three groups of metrics: Peak Signal-to-Noise Ratio (PSNR) and Structural Similarity Index (SSIM) for pixel-wise fidelity, the Learned Perceptual Image Patch Similarity (LPIPS) for perceptual quality, and an adjacent-slice total-variation ratio that directly measures inter-slice consistency. PSNR measures the ratio between the maximum possible pixel value squared and the Mean Squared Error (MSE) between the reconstructed and reference images, providing a quantitative measure of reconstruction quality. SSIM, on the other hand, assesses the visual impact of structural information in the image, making it a more perceptually relevant metric. Because generative reconstruction methods can score well on PSNR/SSIM while still hallucinating implausible texture, or score worse on them while looking diagnostically more natural, we complement both with LPIPS, and we further check for a specific structural failure mode: inter-slice inconsistency along the axis our 2D model processes, with a dedicated TV-ratio metric.

For the low-resolution CBCT results, including the dental, spine, and walnut datasets, we follow the exact evaluation protocol used in~\cite{liu_geometry-aware_2025} to ensure fair comparison with recent state-of-the-art methods.
This includes a predefined clamping range of values for PSNR and SSIM computation. For the dental, spine, and walnut datasets, respectively, we use the following ranges: dental: $[0.0, 0.090]$, spine: $[0.0, 0.051]$, walnut: $[0.0, 0.084]$ (unit-less attenuation coefficients). For SSIM we use an average of the 2D SSIM computed slice-by-slice along the three orthogonal axes (axial, coronal, sagittal) to better capture 3D structural similarity. PSNR is computed in 3D directly.

\paragraph{Perceptual quality (LPIPS)}
PSNR and SSIM are computed directly on attenuation values and are not designed to penalize plausible-but-incorrect texture, a failure mode specific to generative reconstruction methods. We therefore complement them with the Learned Perceptual Image Patch Similarity (LPIPS)~\cite{zhang2018unreasonable}, computed slice-by-slice as the distance between deep-network feature activations of the reconstructed and reference images and averaged over the volume (lower is better). Because posterior averaging (e.g.\ $\mu$(CDPA)) is by construction the MSE-optimal, pixel-wise estimator, LPIPS lets us quantify the realism/accuracy trade-off it introduces relative to a single posterior sample, rather than relying on a PSNR/SSIM gap alone. We report LPIPS for the high-resolution walnut experiments, alongside the explicit-regularization comparison (Sec.~\ref{sec:inter_slice_consistency_hr}, Table~\ref{tab:hr_full_volume}).

\paragraph{Inter-slice consistency (TV ratio)}
Our 2D diffusion model processes slices independently along the axial axis and has no explicit mechanism to enforce consistency between neighbouring axial slices, unlike the coronal and sagittal directions, which it never processes directly; volume-averaged PSNR, SSIM and LPIPS cannot expose this asymmetry. We therefore measure inter-slice consistency directly with an adjacent-slice total-variation (TV) ratio. For ground-truth volume $\xx_0$ and reconstruction $\hat{\xx}_0$, and for each axis $a\in\{\text{axial},\text{coronal},\text{sagittal}\}$, let $\text{TV}_a(V)$ be the mean absolute difference between adjacent slices of $V$ along axis $a$. We define
\begin{equation}
\text{tv\_ratio}_a = \frac{\text{TV}_a(\hat{\xx}_0)}{\text{TV}_a(\xx_0)},
\label{eq:tv-ratio}
\end{equation}
so that $\text{tv\_ratio}_a\approx 1$ means the reconstruction varies from slice to slice along axis $a$ by the same amount as the ground truth, while $\text{tv\_ratio}_a>1$ indicates spurious jitter beyond what the anatomy itself produces. The two off-axis ratios serve as a control for a method's overall sharpness or blur (a uniformly blurry method scores below 1 on \emph{all three} axes), which lets us isolate the axis-specific pathology as
\begin{equation}
\text{tv\_excess} = \text{tv\_ratio}_{\text{axial}} - \tfrac{1}{2}\big(\text{tv\_ratio}_{\text{coronal}}+\text{tv\_ratio}_{\text{sagittal}}\big).
\label{eq:tv-excess}
\end{equation}
$\text{tv\_excess}>0$ specifically flags the axial-direction jitter characteristic of slice-by-slice 2D processing, robust to a method's general smoothness level. We report $\text{tv\_ratio}_{\text{axial}}$ and $\text{tv\_excess}$ throughout Sec.~\ref{sec:results}.

For our high-resolution CBCT results on the walnut dataset, we compute PSNR and SSIM with the same clamping range as in the low-resolution setting using the full volume. However, when comparing with S-STAR Net~\cite{zhang_sliding_2025}, we compute PSNR and SSIM on the central crop of size $360 \times 360 \times 150$ to match their evaluation protocol.

\section{Results} \label{sec:results}

\subsection{Low-resolution CBCT results}

Table~\ref{tab:results_comparison} presents a comprehensive quantitative comparison of our proposed methods against several state-of-the-art baselines across three CBCT datasets with volume size $256^3$: Dental, Spine, and Walnut. The results are reported in terms of Peak Signal-to-Noise Ratio (PSNR) and Structural Similarity Index (SSIM), with values presented as mean ± standard deviation.

Our baseline diffusion model (DPA) shows decent performance, but struggles to surpass some of the recent supervised CNN-based methods such as GAAL~\cite{liu_geometry-aware_2025}, as already noticed by the authors of the GAAL paper. However, when we introduce conditioning on the FDK reconstruction (CDPA), we observe a significant performance boost across all datasets, achieving state-of-the-art results and improving 3D consistency.
Furthermore, when considering the mean of multiple posterior samples with $\mu$(CDPA), we achieve the best PSNR scores across all datasets, strongly suggesting that the posterior distribution is unimodal and that averaging samples effectively reduces noise and artifacts. This mean posterior estimation also helps the baseline DPA method achieve competitive results, although it is still lagging behind CDPA.

Interestingly, our FDK-denoising U-Net model with slice conditioning also achieves competitive results, particularly in SSIM, highlighting the potential of simpler architectures when combined with data-consistency enforcement. It matches or surpasses all previous methods in both PSNR and SSIM, and we can notice that the use of fine-tuning (FDK-denoiser + FT) further improves PSNR results almost to the point of CDPA, this fine-tuning however, might decrease SSIM results, which suggests that it might be overfitting to the MSE loss and losing some of the perceptual quality of the reconstruction.

$\mu$(CDPA) is, by construction, an approximation of the posterior mean, i.e.\ the Bayes-optimal estimator, which explains why it wins on PSNR essentially by definition. Standard bias-variance reasoning predicts this should come at a cost in perceptual/structural quality: averaging several plausible textures is pixel-wise closer to the ground truth, but smoother than any single sample, which SSIM's local structure term can penalize. Surprisingly, this expected trade-off largely fails to materialize: on Dental, $\mu$(CDPA) is also the best method on SSIM (0.921 vs.\ 0.901 for FDK-denoiser), and on Spine it is also the best method on SSIM (0.952 vs.\ 0.950 for FDK-denoiser), i.e.,\ $\mu$(CDPA) improves on both fidelity and structural similarity simultaneously, rather than trading one for the other.
On Walnut at 20 views, the most severely underdetermined setting in the paper (Sec.~\ref{sec:datasets}), $\mu$(CDPA) clearly trails on SSIM (0.813 vs.\ 0.888 for FDK-denoiser and 0.889 for GAAL), though there is still a clear increase in both PSNR and SSIM when averaging over posterior samples.

To further illustrate the advantage of our slice conditioning and the fine-tuning step, we conducted an ablation study discussed in Section~\ref{sec:ablation_study} (although at higher resolution). The results confirm that both components significantly contribute to the overall performance, with slice conditioning providing substantial gains and fine-tuning further enhancing the results to nearly match our diffusion-based methods. Without these components, the U-Net architecture alone falls quite short of state-of-the-art performance.

\subsection{Low-resolution qualitative results}

\begin{figure}[tbp]
\centering

\includegraphics[width=\textwidth]{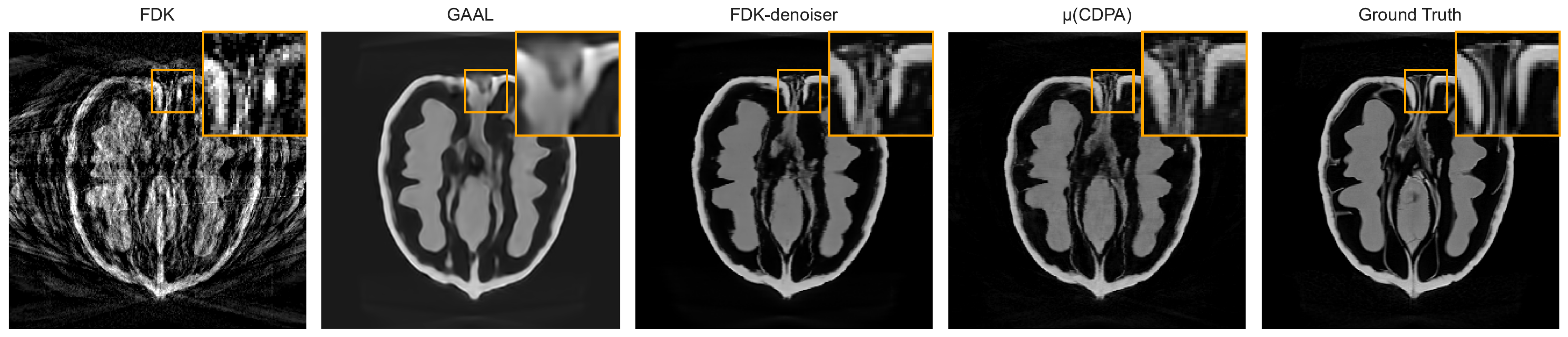}
\includegraphics[width=\textwidth]{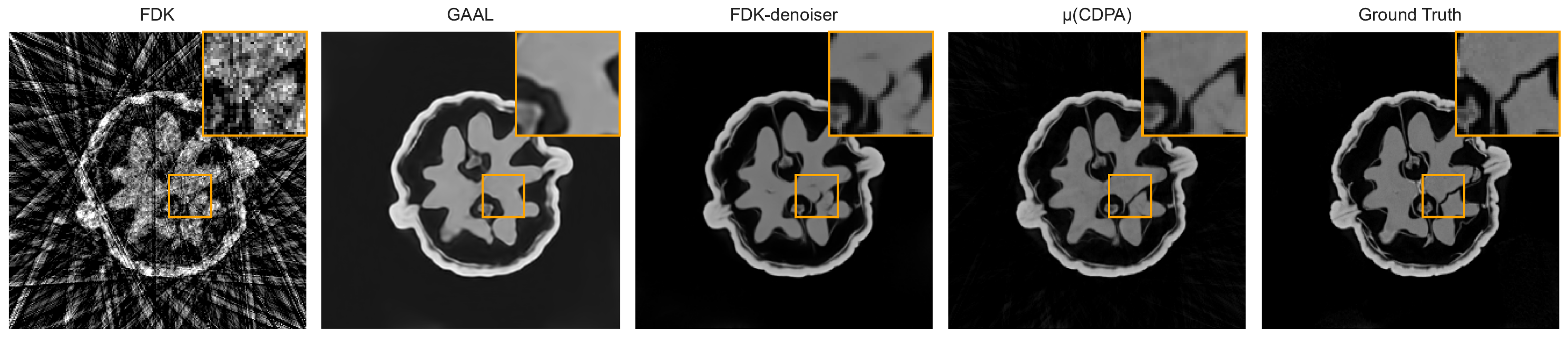}
\includegraphics[width=\textwidth]{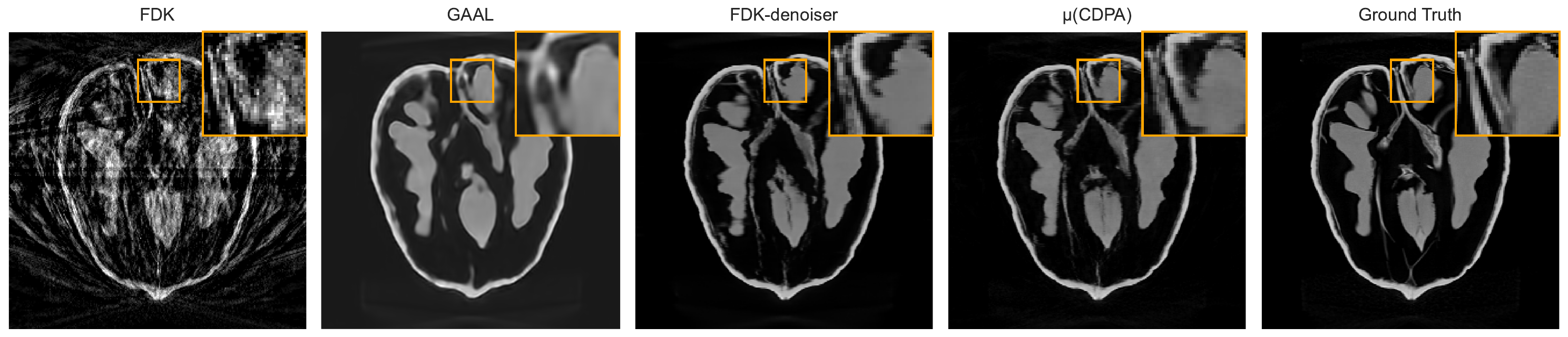}
\caption{Slices (coronal, axial and sagittal from top to bottom) of the walnut reconstruction results with resolution $256^3$ from the test dataset using 20 views uniformly spaced from the 1200 available views from the middle scan.}
\label{fig:walnut_results}
\end{figure}

\begin{figure}[tbp]
\centering
\includegraphics[width=\textwidth]{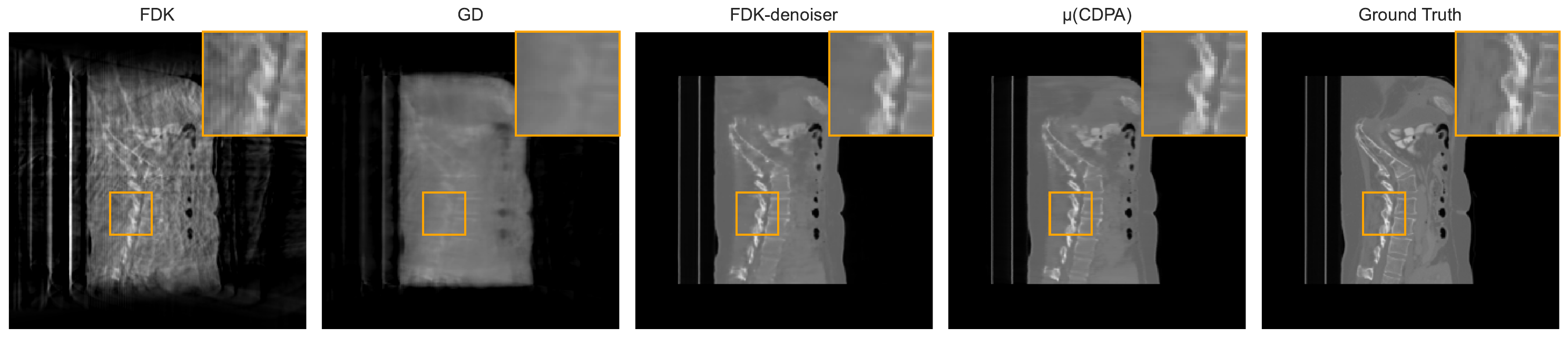}
\includegraphics[width=\textwidth]{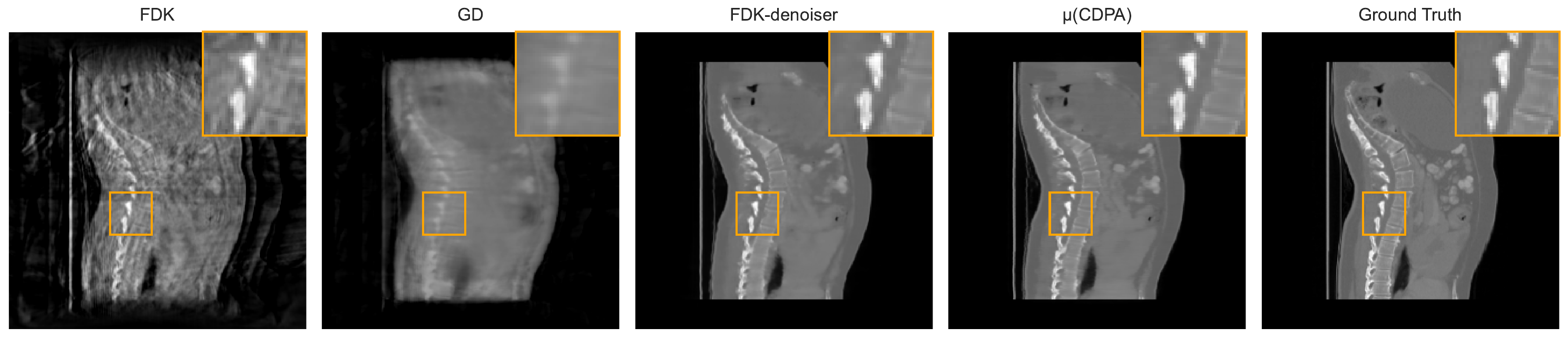}
\caption{Coronal slices of the spine reconstruction results with resolution $256^3$ from the test dataset using 20 views uniformly spaced from 0 to 360 degrees.}
\label{fig:spine_results}
\end{figure}

\begin{figure}[tbp]
\centering
\includegraphics[width=\textwidth]{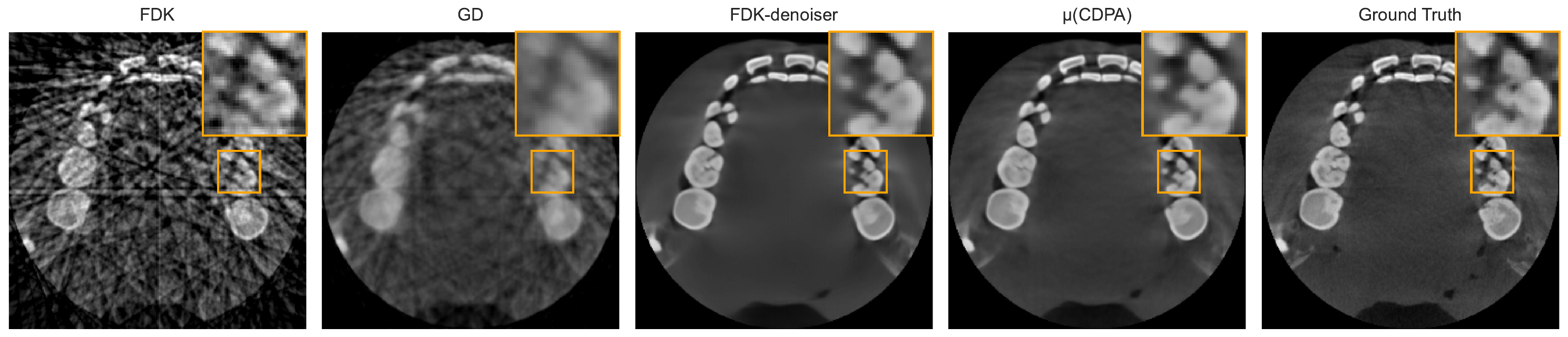}
\includegraphics[width=\textwidth]{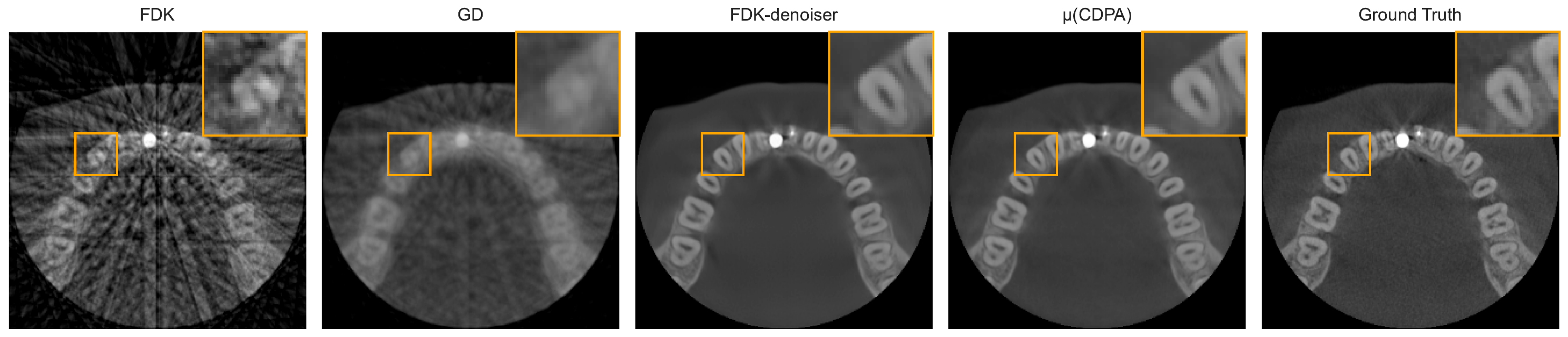}
\caption{Axial slices of the dental reconstruction results with resolution $256^3$ from the test dataset using 20 views uniformly spaced from 0 to 360 degrees.}
\label{fig:dental_results}
\end{figure}

Figures~\ref{fig:walnut_results},~\ref{fig:spine_results} and~\ref{fig:dental_results} present qualitative reconstruction results for the Walnut, Spine and Dental datasets, respectively, at a resolution of $256^3$ voxels using 20 views. These figures show the visual quality of our proposed methods compared to the FDK baseline, GAAL~\cite{liu_geometry-aware_2025} and gradient descent, minimizing data-consistency loss.
We can see that both conditional diffusion and FDK-denoising methods significantly reduce artifacts and enhance structural details compared to FDK and gradient descent, leading to clearer and more accurate reconstructions. 
The improvements of our methods are particularly evident in regions with complex anatomical structures, where we effectively recover fine details that are often lost by other methods.
We observe that hallucinations still exist in this high-sparsity regime because the information is simply not present in the available measurements. However, $\mu$(CDPA) ameliorates this effect by taking the average over a few samples, artifacts hallucinated by a single reconstruction are voted out, leaving a sort of ``majority voting'' solution to the inverse problem. Noticeably, this average reconstruction is sharp, opposite to the usual average reconstruction found in models such as variational autoencoders (VAEs). This suggests a clear unimodal posterior distribution where our mean reconstruction lies closer to the ground truth.

For the walnut data based on real measurements, we can see that the coronal and sagittal views (top and bottom rows in Figure~\ref{fig:walnut_results}) still exhibit some inter-slice inconsistencies, compared with the axial view where the diffusion process is carried out. We can also see that, overall, GAAL produces results with substantially less detail and sharpness. However, producing less detailed reconstructions also means that GAAL can avoid some inter-slice inconsistencies.

Compared to the FDK-denoiser, we can see how diffusion models always produce detailed realistic structures, even if they are not always correct, while FDK-denoising models sometimes produce more unnatural structures, as can be seen in the walnut reconstructions in Figure~\ref{fig:walnut_results}.

\subsection{High-resolution CBCT results vs number of views}
Figure~\ref{fig:high_resolution_results} presents the reconstruction quality (PSNR and SSIM) of our methods on the high-resolution walnut dataset with $501^3$ voxels using varying numbers of views from the high scan. For each sparsity level, we selected a near-uniform subset of views from the 1200 available views in the high scan. We compare our methods against the ground-truth reconstruction using 3600 views from all three scans (high/middle/low).
With 20 views, our models already achieve significantly better PSNR and SSIM than the FDK baseline using all 1200 views from the high scan, and with 40 views they outperform the gradient descent reconstruction using all 1200 views from the high scan, highlighting the effectiveness of our methods in extremely sparse-view scenarios. CDPA with mean posterior averaging, $\mu$(CDPA), provides the best performance across all view counts, with a consistent improvement of approximately 1~dB of PSNR over standalone CDPA, demonstrating its ability to effectively leverage both learned priors and data consistency. Similarly, FDK-denoising with fine-tuning (+ FT) achieves competitive results, closely matching CDPA, especially at higher view counts.

\subsection{High-resolution results at $N=30$ views and inter-slice consistency}\label{sec:inter_slice_consistency_hr}
For the specific case of 30 views, we fine-tuned our model with samples at this specific sparsity. We acknowledge that this fine-tuning step provides a significant advantage over the model trained for varying numbers of views, as shown in Figure~\ref{fig:high_resolution_results}, where we compare metrics for the whole generated volume.

We compare our results for $N=30$ with S-STAR Net~\cite{zhang_sliding_2025} in Table~\ref{tab:walnut_metrics}, showing that our method slightly exceeds S-STAR Net and outperforms other methods in both SSIM and PSNR, demonstrating that our 2D approach scales to the performance of state-of-the-art 3D approaches.
Recall that these numbers are not directly comparable with other tables as they are computed only within a central crop; thus, while the SSIM values remain similar, the PSNR values differ from those of the entire volume, as almost no background is considered.

\begin{table}[ht]
\centering
\caption{\small Walnut high-resolution results for the center $360\times360\times150$ crops conditioned on 30 views. Results with a star are taken from \cite{zhang_sliding_2025}.}
\label{tab:walnut_metrics}
\begin{tabular}{lcc}
\toprule
\textbf{Algorithm} & \textbf{SSIM$\uparrow$} & \textbf{PSNR$\uparrow$} \\
\hline
U-former$^*$   & $0.87 \pm 0.01$ & $24.94 \pm 1.37$ \\
PTNet3D$^*$    & $0.82 \pm 0.01$ & $22.09 \pm 0.36$ \\
Pix2Pix3D$^*$  & $0.85 \pm 0.01$ & $23.74 \pm 0.56$ \\
S-STAR Net$^*$ & $0.90 \pm 0.01$ & $27.17 \pm 0.95$ \\
\hline
$\mu$(CDPA) (ours) & $\mathbf{0.91 \pm 0.01}$ & $\mathbf{27.35 \pm 1.47}$ \\
\bottomrule
\end{tabular}
\end{table}

Table~\ref{tab:walnut_metrics} follows S-STAR Net's protocol of evaluating only a central $360{\times}360{\times}150$ crop, since that baseline cannot process the full volume. This means our defensible advantage there is not the small margin in the table itself, but that our 2D approach reaches this quality on the \emph{entire} $501^3$ volume with the same $N=30$-view fine-tuned model, which we report in Table~\ref{tab:hr_full_volume} using the same PSNR, SSIM, LPIPS~\cite{zhang2018unreasonable} and TV-ratio metrics defined in Sec.~\ref{sec:metrics}.

This also lets us test the inter-slice-consistency claim on real CBCT measurements at a view count where every method has enough information to produce a reasonable reconstruction. The conditioning benefit is clear, DPA's axial TV excess lands at +1.00 and CDPA's at +0.55, and $\mu$(DPA)'s excess of +0.50 is cut to $\mu$(CDPA)'s +0.14. PSNR and SSIM improve substantially for every method between 20 and 30 views, and the CDPA-over-DPA margin widens rather than narrows ($\mu$(CDPA) reaches 34.57~dB/0.906 SSIM vs.\ $\mu$(DPA)'s 29.56~dB/0.747), consistent with conditioning extracting more benefit from FDK reconstructions that themselves improve with more views.

\begin{table*}[tbp]
\centering
\caption{\small Walnut, full $501^3$ volume, $N=30$ views (fine-tuned model), $n=5$ test volumes. $\text{tv\_excess}$ as defined in Eq.~\ref{eq:tv-excess}. DPA+ADMM-TV and CDPA+ADMM-TV apply DiffusionMBIR's explicit ADMM+1D-TV guidance (Sec.~\ref{sec:mbir_comparison}) to our pretrained checkpoints; all rows are averaged over the same $n=5$ test volumes.}
\label{tab:hr_full_volume}
\resizebox{\textwidth}{!}{%
\begin{tabular}{lccccc}
\toprule
\textbf{Method} & \textbf{PSNR} & \textbf{SSIM} & $\textbf{tv\_ratio}_{\textbf{axial}}$ & $\textbf{tv\_excess}$ & \textbf{LPIPS} \\
\hline
FDK & 16.27±0.40 & 0.124±0.002 & 11.556 & $-$1.125 & 0.809±0.004 \\
\hline
DPA & 27.63±0.59 & 0.727±0.021 & 3.265 & +1.001 & 0.337±0.024 \\
CDPA & 32.57±0.55 & 0.879±0.008 & 1.509 & +0.551 & 0.107±0.009 \\
DPA+ADMM-TV & 32.04±0.57 & 0.767±0.012 & 1.486 & $-$0.269 & 0.227±0.012 \\
CDPA+ADMM-TV & 32.60±0.66 & 0.772±0.012 & 1.490 & $-$0.273 & 0.228±0.012 \\
FDK-denoiser & 32.04±0.72 & 0.891±0.007 & 0.655 & \textbf{+0.022} & 0.166±0.009 \\
\hline
$\mu$(DPA) & 29.56±0.53 & 0.747±0.019 & 2.634 & +0.495 & 0.304±0.020 \\
$\mu$(DPA+ADMM-TV) & 32.36±0.60 & 0.772±0.012 & 1.465 & $-$0.252 & 0.220±0.012 \\
$\mu$(CDPA+ADMM-TV) & 32.87±0.67 & 0.777±0.011 & 1.471 & $-$0.255 & 0.220±0.011 \\
$\mu$(CDPA) & \textbf{34.57±0.68} & \textbf{0.906±0.007} & \textbf{1.001} & +0.144 & \textbf{0.092±0.006} \\
FDK-denoiser + FT & 31.11±0.58 & 0.763±0.020 & 2.160 & +0.232 & 0.278±0.019 \\
\bottomrule
\end{tabular}%
}
\end{table*}

\begin{figure}[tbp]
\centering
\includegraphics[width=\textwidth]{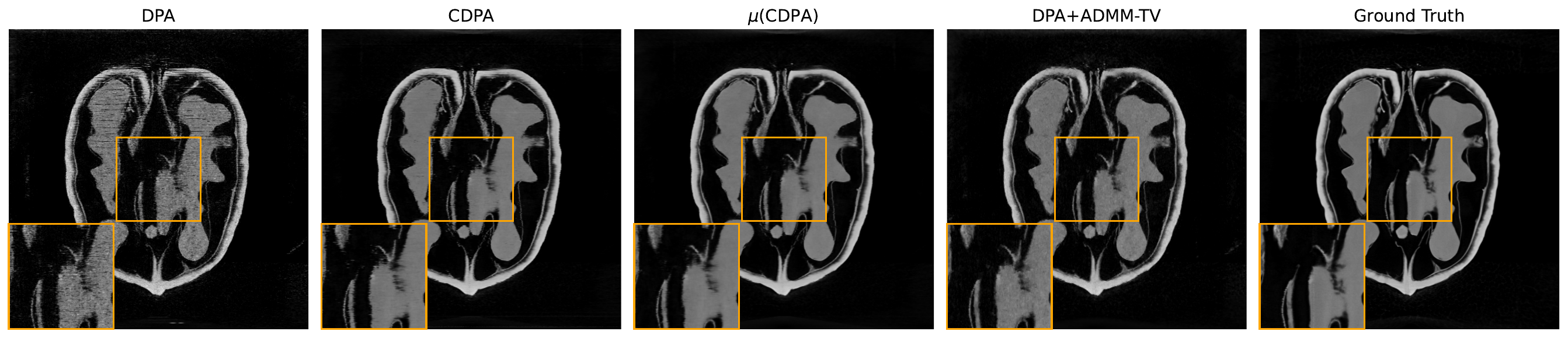} 
\includegraphics[width=\textwidth]{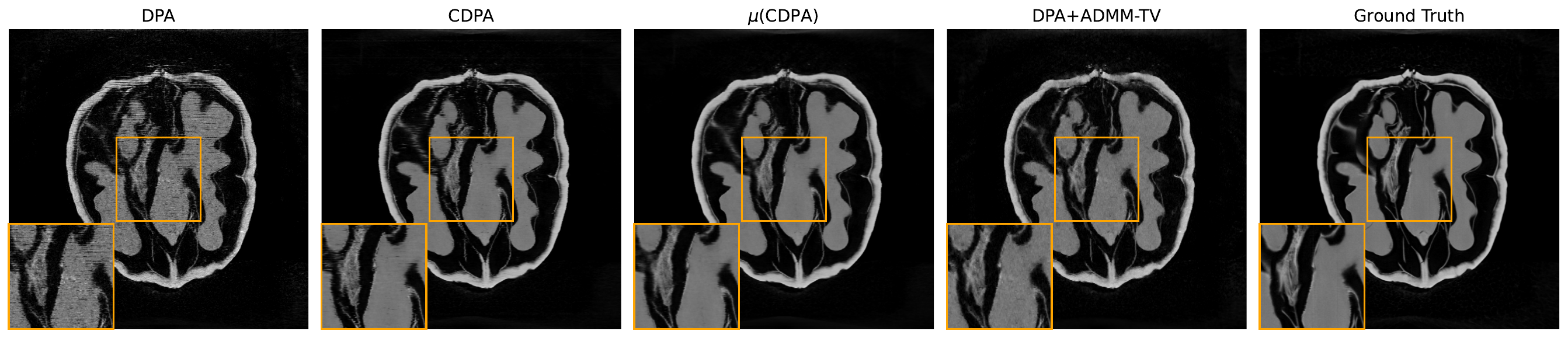}
\caption{Slices (coronal and sagittal from top to bottom) of the walnut reconstruction results with resolution $501^3$ from the test dataset using 30 views uniformly spaced from the 1200 available views from the middle scan.}
\label{fig:walnut_results_501}
\end{figure}

\paragraph{FDK-Conditioning vs.\ Explicit Regularization}\label{sec:mbir_comparison}

An alternative to FDK-conditioning for enforcing inter-slice consistency is to impose an explicit hand-crafted regularizer directly on the diffusion sampling process. DiffusionMBIR~\cite{chung2023solving3d} augments a \emph{pretrained, unconditional} 2D diffusion prior with an ADMM data-consistency step that solves
$\min_{\xx}\, \tfrac{1}{2}\|\A\xx-\yy\|_2^2 + \lambda\|D_z\xx\|_1$
at every reverse-diffusion timestep, where $D_z$ is the first-difference operator along the axial axis -- a 1D total-variation penalty coupling adjacent slices, solved via alternating conjugate-gradient (CG) least-squares updates and soft-thresholding. We implement this guidance mechanism as a drop-in replacement for our own ReSample-style gradient-descent guidance (Sec.~\ref{sec:Method}), which lets us apply it to our \emph{existing, pretrained} DPA and CDPA checkpoints with no retraining -- DPA+ADMM-TV is exactly DiffusionMBIR; CDPA+ADMM-TV additionally retains our FDK-conditioning, testing whether the two consistency mechanisms are complementary or redundant.
Parameter $\lambda$ and the ADMM/CG iteration budget have no principled default and must be tuned per acquisition, an expensive per-dataset hyperparameter search CDPA does not require.

\paragraph{Fidelity is close, but that is not the whole story} The DPA+ADMM-TV and CDPA+ADMM-TV rows of Table~\ref{tab:hr_full_volume} show that CDPA remains the strongest single-sample method on both PSNR and, especially, SSIM, even against this properly-tuned explicit-regularization baseline: SSIM trails CDPA by 0.107--0.112, while PSNR is within 0.5~dB.
On PSNR/SSIM alone, the two strategies for restoring 3D consistency look close to interchangeable. However, they are not: the following three paragraphs identify three independent, easily measurable axes on which explicit regularization loses, none of which a pixel-fidelity metric alone would reveal.

\paragraph{Cost 1: axial over-smoothing} Every other row of Table~\ref{tab:hr_full_volume} has a \emph{positive} $\text{tv\_excess}$ -- the axial-specific jitter the metric was designed to detect (Sec.~\ref{sec:inter_slice_consistency_hr}). Both ADMM-TV variants instead show a \emph{negative} $\text{tv\_excess}$ ($-0.269$ and $-0.273$): the ADMM penalty does not merely remove the jitter, it over-corrects, smoothing the axial direction \emph{more} than the anatomy's own in-plane structure would predict. This is the opposite pathology from the one motivating CDPA in the first place, and it is invisible to PSNR/SSIM -- exactly why a metric that measures inter-slice consistency directly, rather than inferring it from a fidelity score, is needed to catch it.

\paragraph{Cost 2: a collapsed posterior} Our results show that averaging posterior samples improves both fidelity and structural similarity for DPA/CDPA, which we attributed to a near-unimodal posterior: in Table~\ref{tab:hr_full_volume}, $\mu$(DPA) and $\mu$(CDPA) gain +1.93~dB and +2.00~dB PSNR respectively over their single-sample counterparts, with a substantial SSIM gain for CDPA ($0.879\to0.906$). The $\mu$(DPA+ADMM-TV) and $\mu$(CDPA+ADMM-TV) rows test whether this benefit survives once the explicit, deterministic ADMM constraint is added to the same sampling loop ($n=10$ posterior samples, same $n=5$ test volumes and hyperparameters as the single-sample ADMM-TV rows). For both variants the gain collapses to well under a sixth of DPA/CDPA's own gain -- $+0.32$ and $+0.27$~dB, with a correspondingly modest SSIM gain ($0.767\to0.772$ and $0.772\to0.777$) -- regardless of conditioning. This is a genuine posterior-diversity effect rather than a metric artifact: because the CG-solved ADMM update is a deterministic function of the fixed measurement $\yy$, it pulls every draw toward nearly the same fixed point regardless of the diffusion model's own stochastic noise, leaving little independent sample-to-sample variation for Monte Carlo averaging to exploit. This collapse is not a conditioning artifact -- it is essentially identical with and without FDK-conditioning -- so it is a property of the explicit regularizer itself, not of its interaction with conditioning.

\paragraph{Cost 3: substantially worse perceptual quality} Table~\ref{tab:hr_full_volume}'s LPIPS column tells a third, independent story. Both ADMM-TV variants score $0.227$--$0.228$ -- more than double CDPA's $0.107$ and more than double $\mu$(CDPA)'s $0.092$ -- despite reaching comparable PSNR and SSIM to CDPA. Enforcing 3D consistency through an explicit spatial penalty evidently costs perceptual realism that pixel-wise fidelity metrics cannot see, consistent with the over-smoothing already identified via $\text{tv\_excess}$ (Cost 1) and independent of the posterior-collapse argument (Cost 2), which concerns sample diversity rather than any single sample's quality. Posterior averaging, which meaningfully reduces CDPA's LPIPS ($0.107\to0.092$), barely moves it for either ADMM-TV variant ($0.227\to0.220$ and $0.228\to0.220$) -- a third symptom of Cost 2's collapsed diversity: with little independent variation left to average over, there is little perceptual benefit left to gain from averaging either.

\subsection{High-resolution qualitative results}

\begin{figure}[tbp]
\centering
\includegraphics[width=\textwidth]{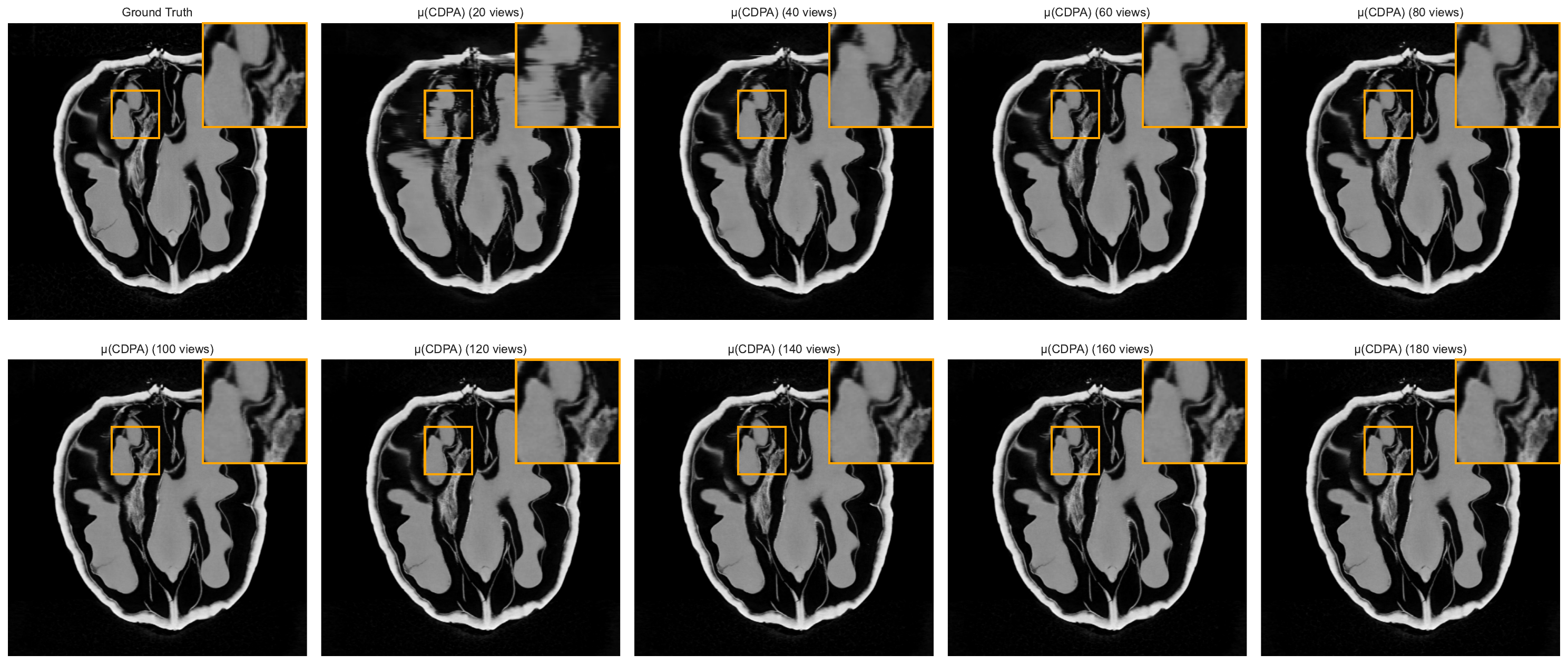}
\caption{Axial slices of a walnut reconstruction using $\mu$(CDPA) at resolution $501^3$ for the first walnut in the test dataset using progressively more views. Each view count is spaced uniformly from the 1200 available views from the high scan. The 3D consistency improves progressively as well; with only 20 views one can see inter-slice inconsistencies, but shortly after they become imperceptible.}
\label{fig:hr_walnut_results}
\end{figure}

\begin{figure}[htbp]
\centering
\includegraphics[width=0.49\textwidth]{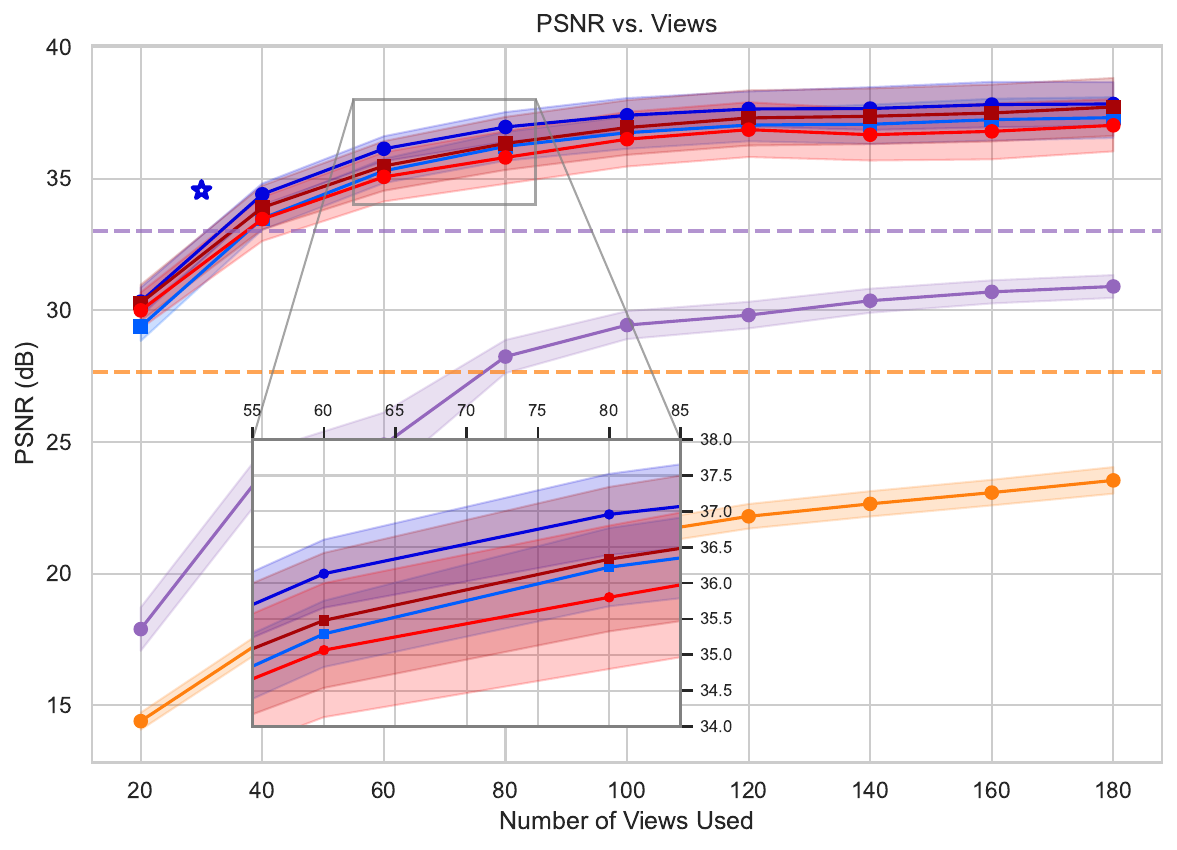}
\includegraphics[width=0.49\textwidth]{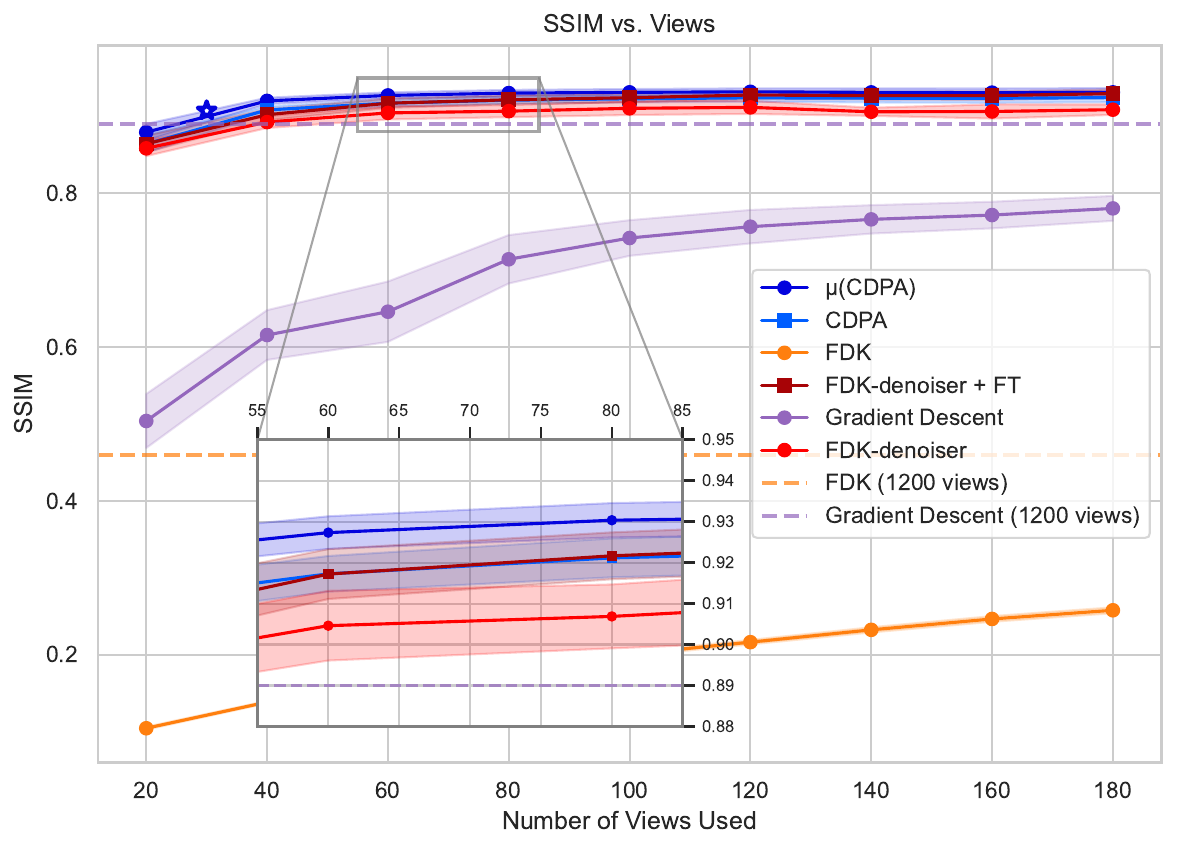}
\caption{PSNR and SSIM vs number of views for the high-resolution walnut dataset at resolution $501^3$ using the high-scan setting. The model was jointly trained for all sparsities. Dotted lines represent the baseline reconstruction using FDK and GD with all 1200 views from this high scan. All metrics are computed by comparing with the ground-truth reconstruction using 3600 views from all three high/middle/low scans. The star represents the result of the fine-tuned model for 30 views, showing the improvement that fine-tuning on a specific sparsity provides.}
\label{fig:high_resolution_results}
\end{figure}

Figure \ref{fig:hr_walnut_results} presents qualitative reconstruction results for the high-resolution walnut dataset with $501^3$ voxels using between 20 and 180 views from the high scan. The results show a progressive improvement in the quality of the reconstruction as the number of views increases. With only 20 views, the reconstructions exhibit still significant hallucinations, and there is not enough inter-slice consistency, although the quality of the reconstruction remains high. However, as the number of views increases, our method gets closer and closer to the ground truth, with almost no discernible difference at 100 views compared to the ground truth that uses 3600 views from all three scans (high, middle and low scans).
Already with 40 views (or 30 in the fine-tuned model), inter-slice inconsistencies mostly disappear, which is supported by the largest jump in SSIM and PSNR in Figure~\ref{fig:high_resolution_results} after going above 20 views.

\subsection{Computational Efficiency}

In terms of speed, Figure~\ref{fig:ablation_unet} compares the running time of our different methods on the CBCT datasets with size $256^3$. Conditional diffusion requires fewer data-consistency steps than its unconditional counterpart due to the information already present in the FDK reconstruction. While these numbers may vary with the size of the GPU and how many slices are produced in one single batch, there is no denying that diffusion-based methods are significantly more computationally intensive than FDK-denoising approaches. Notably, FDK-denoising achieves performance comparable to diffusion-based models but with up to 18$\times$ faster runtime, making it a practical choice for real-world applications. However, neither method has been optimized for speed, and deployment ready software is not the objective of this paper.

\subsection{Ablation Studies}\label{sec:ablation_study}
\begin{figure}[htbp]
\centering
\includegraphics[width=0.34\textwidth]{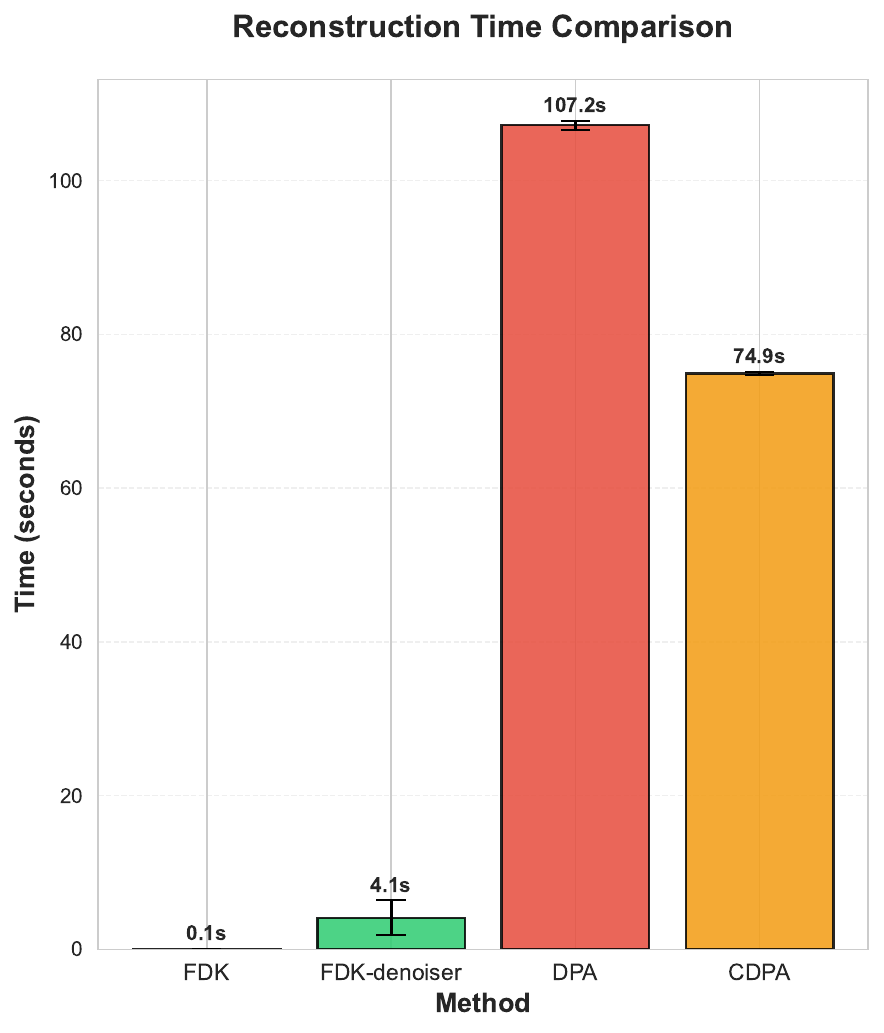}
\includegraphics[width=0.54\textwidth]{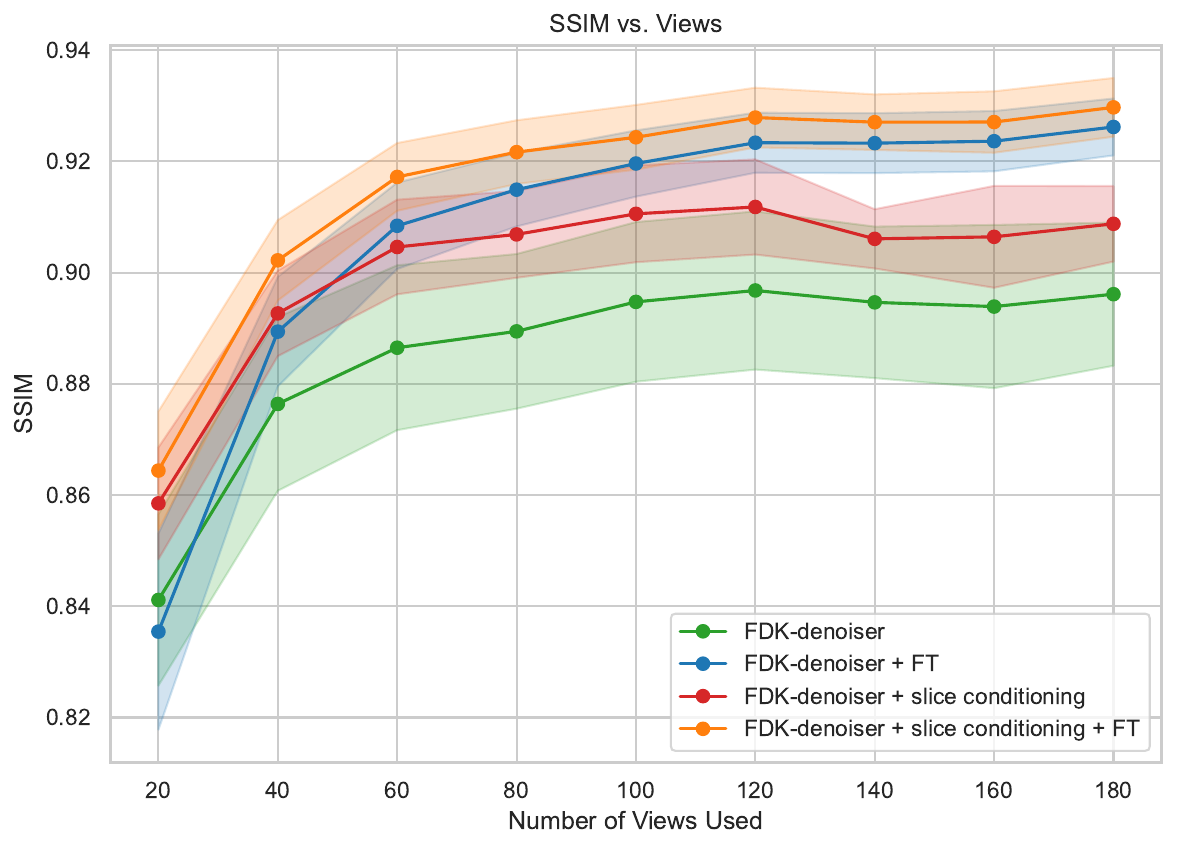}
\caption{\textbf{Left:} Running time comparison of different methods on the CBCT datasets with size $256^3$. Conditional diffusion requires fewer data-consistency steps than its unconditional counterpart due to the information already provided by the conditioning FDK reconstruction. FDK-denoising with fine-tuning is significantly faster than diffusion-based methods while achieving comparable performance. \textbf{Right:} Ablation study of slice conditioning and data-consistency fine-tuning (FT) on the Walnut dataset at full resolution of $501^3$ voxels. A modern U-Net architecture on its own is competitive but is not state-of-the-art. The addition of slice conditioning significantly improves performance, and data-consistency fine-tuning (FT) further boosts the results to nearly match our proposed diffusion-based methods.}
\label{fig:ablation_unet}
\end{figure}

Figure~\ref{fig:ablation_unet} (right) presents an ablation study of slice conditioning and data-consistency fine-tuning (FT) on the Walnut dataset at full resolution of $501^3$ voxels.
The results confirm that both components significantly contribute to the overall performance, with slice conditioning providing substantial gains and fine-tuning further enhancing the results to nearly match our diffusion-based methods.
Interestingly, FT has a detrimental effect when working with only 20 views and without slice conditioning, but becomes beneficial when combined with slice conditioning at this high sparsity. This suggests that the fine-tuning step becomes beneficial when the original denoising network is already producing reasonable reconstructions, but is susceptible to overfitting in the ultra-sparse regime.
Moreover, fine-tuning becomes increasingly beneficial in all settings as the number of views increases, matching the intuition that data-consistency becomes the dominant source of information as more measurements are available.

Without these components, the U-Net architecture alone falls quite short of state-of-the-art performance. The improvement carries over to different sparsities, though it is more dramatic at higher sparsities (fewer views).
These two components are enough to make FDK-denoising competitive with diffusion-based models, and redefine state-of-the-art, while being significantly faster.

\section{Uncertainty Quantification}\label{sec:uncertainty_quantification}

\begin{figure}[htbp]
\centering
\includegraphics[width=\textwidth]{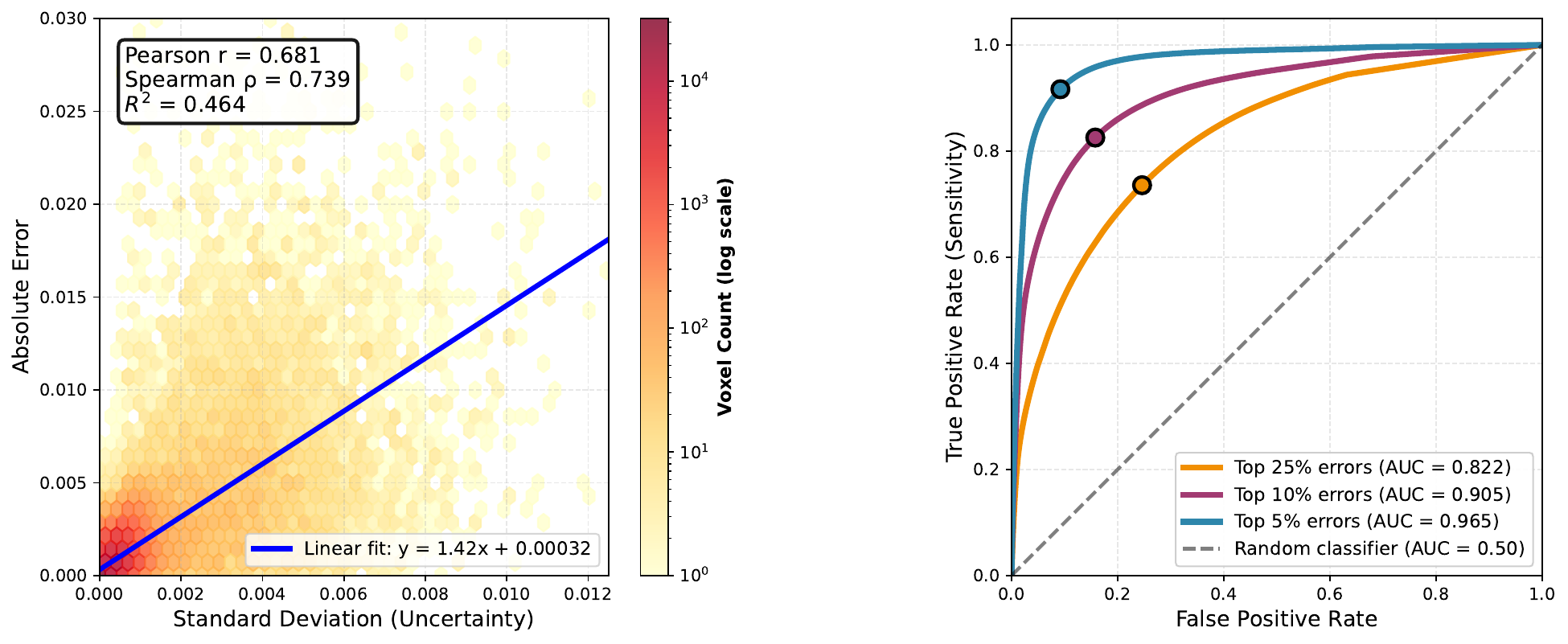}
\includegraphics[width=\textwidth]{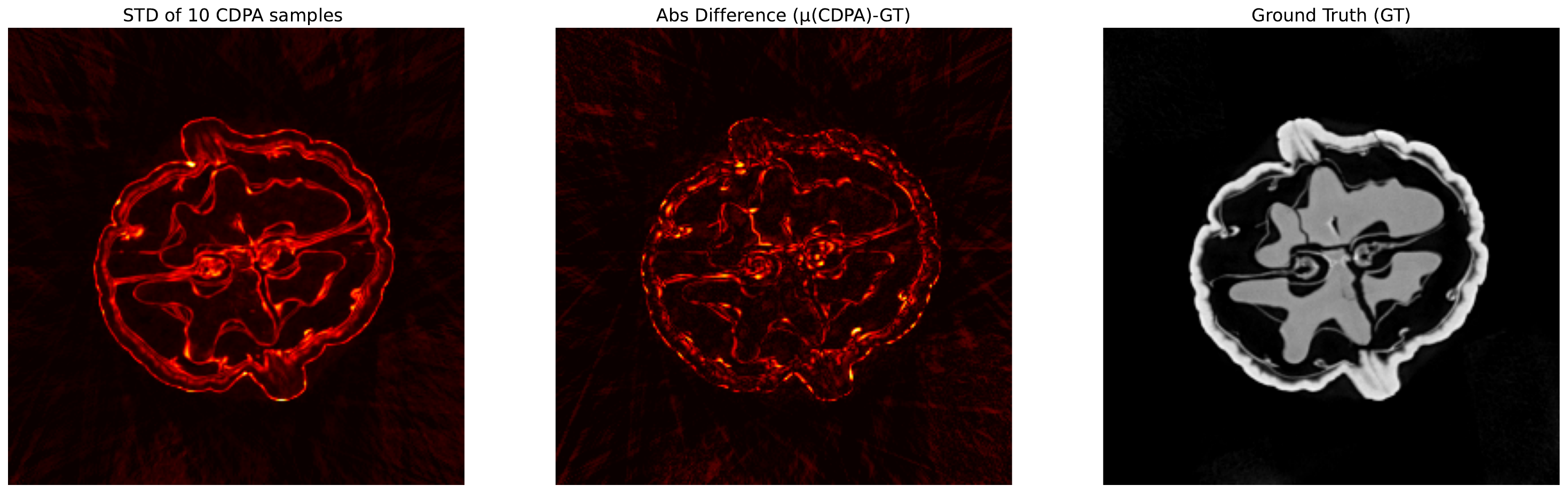}
\caption{
\textbf{(A) Top left:} Correlation between standard deviation (STD) of diffusion samples for the $256^3$ walnut dataset and reconstruction error using 20 projections. The hexbin density plot shows strong positive correlation (Pearson $r = 0.680$, Spearman $\rho = 0.735$, $R^2 = 0.462$), with linear fit $\text{Error} \approx 1.42 \times \text{STD} + 0.0003$. STD explains 46.2\% of error variance, demonstrating its utility as an uncertainty indicator.
\textbf{(B) Top right:} Receiver Operating Characteristic (ROC) curves for identifying high-error regions using STD thresholds. The method achieves excellent discrimination with AUC values of 0.820, 0.902, and 0.961 for detecting the top 25\%, 10\%, and 5\% highest-error voxels, respectively. Optimal operating points (circles) demonstrate practical trade-offs between sensitivity and specificity. At the optimal threshold for top 5\% errors, the method achieves 91.3\% sensitivity with only 9.2\% false positive rate.
\textbf{(C) Bottom:} Spatial comparison of a representative axial slice showing ground truth, uncertainty map (STD), and absolute error. High-uncertainty regions (bright areas in STD map) spatially correspond to high-error regions, validating STD as a reliable spatial predictor of reconstruction quality.}
\label{fig:uncertainty_analysis}
\end{figure}

Uncertainty quantification is an important aspect in medical imaging applications, where understanding the confidence of the reconstruction can inform clinical decision-making and guide further imaging or interventions. We explore the use of $\mu$(CDPA) to quantify uncertainty in the reconstructions, by computing multiple samples from the posterior distribution.
More specifically, we sample several times from $p(\xx | \yy)$ using $\mu$(CDPA) and consider the mean of these samples, as well as their pixel-wise standard deviation (STD).
To assess the reliability of our uncertainty estimates, we analyzed the relationship between the STD of diffusion samples and reconstruction error against ground truth. Figure~\ref{fig:uncertainty_analysis} presents a comprehensive uncertainty quantification analysis with a qualitative comparison of a representative axial slice, a correlation analysis between STD and absolute error, and a Receiver Operating Characteristic (ROC) analysis for identifying high-error regions using STD thresholds.

\paragraph{Correlation Analysis}
The STD exhibits a positive correlation with the absolute reconstruction error (Pearson $r = 0.680$, Spearman $\rho = 0.735$, both $p < 10^{-10}$), explaining 46.2\% of error variance ($R^2 = 0.462$) (Fig.~\ref{fig:uncertainty_analysis}A). The linear relationship ($\text{Error} \approx 1.42 \times \text{STD} + 0.0003$) indicates that regions with higher sampling variability consistently exhibit larger reconstruction errors. The higher Spearman correlation suggests a monotonic relationship that slightly deviates from perfect linearity, typical of heterogeneous structures.

\paragraph{Discriminative Power for Quality Control}
To evaluate the practical utility for identifying problematic reconstructions, we performed ROC analysis treating STD (normalized) as a predictor probability score for high-error regions (Fig.~\ref{fig:uncertainty_analysis}B). The method achieves strong discrimination across different error thresholds: AUC = 0.820 for top 25\% errors, AUC = 0.902 for top 10\% errors, and AUC = 0.961 for top 5\% errors. Notably, the highest AUC for the most severe errors indicates that STD is particularly reliable for identifying critical reconstruction failures.

\paragraph{Discussion} Our analysis shows promise of using the Diffusion posterior for uncertainty quantification. The results represented in Fig.~\ref{fig:uncertainty_analysis} demonstrate clearly that the standard deviation of the Diffusion posterior is a strong predictor of the reconstruction error and, as such, can aid in the interpretation of the reconstruction. However, reliable uncertainty quantification for tomographic reconstruction remains a challenging topic. Directly relying on Diffusion posterior sampling does not alleviate the possibility of misspecification and hallucinations, and Diffusion posterior alignment does not sample the exact Bayesian posterior, c.f.~\cite{moroy2026posterior}. Moreover, the Diffusion posterior samples do not come with formal asymptotic guarantees, such as, for example, Bootstrap based methods \cite{pereyra2024equivariant} or the Laplace approximation \cite{antoran2023uncertainty}, which we therefore expect to perform better in the dense, noiseless data limit. Prior work has further investigated Bayesian models, which typically rely on approximations such as Markov-Chain Monte Carlo (MCMC), e.g.~\cite{durmus2018efficient,pereyra2020accelerating}. Finally, recent work explores the use of sequential likelihood mixing to obtain formal confidence guarantees for neural network and diffusion-based reconstructions \cite{gatzner2026principled}, however these are defined for a level set of the likelihood function, and not by the STD map alone. We leave further exploration of this topic for future work.

\section{Conclusions}\label{sec:conclusions}

In this work, we have presented two state-of-the-art reconstruction methods for sparse-view CBCT reconstruction: Conditional Diffusion Posterior Alignment (CDPA) and a new perspective on the FBP/FDK-denoiser models that ensures data consistency at inference time using fine-tuning.

CDPA leverages the power of diffusion models while incorporating conditioning on the FDK reconstruction to enhance performance and 3D consistency, and the guidance provided by the data-consistency through the diffusion steps. That is, we are effectively combining two techniques to sample from the posterior distribution and show that this combination is synergistic and leads to significant performance improvements. In addition, we noticed that averaging multiple posterior samples further boosts performance, suggesting that the posterior distribution is unimodal and that averaging helps reduce noise and artifacts. This is only the case if the conditioning is strong enough, as is the case in several inverse problems. In addition, computing several samples from the posterior distribution allows us to quantify uncertainty in the reconstructions, which is an important aspect in medical imaging applications.

FDK-conditioning is, however, only one of two ways to restore 3D consistency to a slice-by-slice 2D diffusion prior that we study; the other imposes an explicit, hand-crafted regularizer directly on the sampling process, as DiffusionMBIR does with an ADMM+total-variation guidance step. We compared the two directly, applying the explicit-regularization mechanism to our own pretrained checkpoints with no retraining (Sec.~\ref{sec:mbir_comparison}), and found that the raw fidelity metrics alone are close to a tie: PSNR is comparable and SSIM trails CDPA by only 0.11 SSIM.
Three further, independently measurable properties break that tie decisively in favor of FDK-conditioning: explicit regularization over-smooths the axial axis rather than merely removing its inconsistency (a negative rather than positive $\text{tv\_excess}$), it reduces the sampler's posterior variance to the point that averaging via the $\mu(\cdot)$ operator loses most of its value, and it produces reconstructions that a perceptual metric (LPIPS) scores as markedly worse. None of these three costs is visible in PSNR or SSIM alone, which is precisely why we measured them directly rather than relying on aggregate fidelity to adjudicate between the two strategies. We conclude that, whenever paired training data is available to learn the conditioning in the first place, it is preferable to hand-crafted spatial regularization even at matched fidelity.

The FDK-denoising approach, on the other hand, combines a U-Net architecture with slice conditioning and a data-consistency fine-tuning step to achieve competitive results with significantly reduced computational cost compared to diffusion-based methods. An ablation of these two components confirms their importance in achieving state-of-the-art performance, with slice conditioning providing substantial gains and fine-tuning further enhancing the results to nearly match our diffusion-based methods. Moreover, the computational efficiency of the FDK-denoising approach makes it a practical choice for real-world applications, especially in clinical settings where time constraints are critical.

Our methods scale to high-resolution CBCT reconstruction, where other approaches have failed to do so. The approach can be further scaled up by using latent diffusion techniques, which would allow to work with even higher resolution volumes, up to Gigavoxels resolution, though data-consistency updates require significant changes to work in latent space~\cite{song2023solving}. However, the limiting factor becomes the data-consistency gradient-descent step that at the moment consumes around 100GB of memory for $\approx 1500^3$ resolution volumes.

While 3D-based diffusion models should further improve performance, especially in terms of 3D consistency, they come at a significant computational cost that is at the moment prohibitive for high-resolution volumes. Moreover, a limited number of training volumes, as it is the case for the walnut dataset, makes it almost impossible to train a 3D-based diffusion model without overfitting, while 2D-based models can be trained with a much smaller number of volumes by leveraging the large number of slices available in each volume.

We believe the use of high-sparsity reconstructions (e.g., 20 views) might be only for benchmarking purposes, real clinical applications would likely use more views to ensure high-fidelity reconstructions, but our results show that as low as 180 views (15\% of the original 1200 views from the high scan and only 5\% from the 3600 total views used for the ground truth) are sufficient to achieve near-perfect reconstructions, significantly reducing radiation dose while maintaining high image quality.

\paragraph{Limitations} The method we presented is supervised and sits at the frontier of what can be achieved with high quality training data. We acknowledge, however, that in many applications such training data is not readily available. Moreover, generative reconstruction can hallucinate structure; the uncertainty map mitigates but does not eliminate this, and no reader study or task-based validation has been performed, so these methods are not yet validated for diagnostic use.

\section*{Acknowledgments}

The authors would like to thank AmirEhsan Khorashadizadeh, Raphael Winkler and all participants of the CHIP project of SDSC for useful discussions. 

\bibliographystyle{plain}
\bibliography{references}

\appendix
\section{Supplementary Material}
\label{sec:supplementary}

\subsection{Full Inter-Slice-Consistency Tables ($256^3$)}
\label{sec:supplementary_tv}
Tables~\ref{tab:tv_ratio} and~\ref{tab:axis_gap} give the full per-method, per-dataset breakdown behind the compacted discussion in Sec.~\ref{sec:inter_slice_consistency_hr}, covering all nine methods on all three $256^3$ datasets rather than only the DPA/CDPA/$\mu$(CDPA) comparison highlighted there.

\begin{table*}[tbp]
\centering
\caption{Adjacent-slice total-variation ratio along the axial axis, $\text{tv\_ratio}_{\text{axial}}$ (Eq.~\ref{eq:tv-ratio}), and its axial excess over the in-plane average, $\text{tv\_excess}$ (Eq.~\ref{eq:tv-excess}), at $256^3$/20 views. Values near 0 for $\text{tv\_excess}$ indicate no axis-specific inconsistency; larger positive values indicate spurious inter-slice jitter specific to the axial direction. Same $n$ per dataset as Table~\ref{tab:results_comparison}.}
\label{tab:tv_ratio}
\resizebox{\textwidth}{!}{%
\begin{tabular}{l|cc|cc|cc}
\toprule
\multirow{2}{*}{\textbf{Method}} & \multicolumn{2}{c|}{\textbf{Dental}} & \multicolumn{2}{c|}{\textbf{Spine}} & \multicolumn{2}{c}{\textbf{Walnut}} \\
\cmidrule(lr){2-3} \cmidrule(lr){4-5} \cmidrule(lr){6-7}
& $\text{tv\_ratio}_{\text{axial}}$ & $\text{tv\_excess}$ & $\text{tv\_ratio}_{\text{axial}}$ & $\text{tv\_excess}$ & $\text{tv\_ratio}_{\text{axial}}$ & $\text{tv\_excess}$ \\
\midrule
FDK & 1.776 & $-$0.482 & 2.838 & $-$0.456 & 5.479 & $-$0.781 \\
GD$_{zero}$ & 0.804 & +0.058 & 1.132 & +0.192 & 2.357 & +0.273 \\
GD$_{FDK}$ & 1.045 & $-$0.090 & 1.995 & +0.377 & 2.715 & +0.275 \\
\midrule
DPA & 1.952 & \textbf{+1.061} & 2.367 & \textbf{+1.369} & 2.560 & +0.490 \\
CDPA & 1.232 & +0.439 & 1.429 & +0.628 & 1.510 & +0.355 \\
FDK-denoiser & 0.645 & $-$0.003 & 0.750 & $-$0.037 & 0.751 & $-$0.027 \\
\midrule
$\mu$(DPA) & 0.990 & +0.268 & 1.206 & +0.352 & 2.321 & +0.380 \\
$\mu$(CDPA) & 0.788 & +0.112 & 0.855 & +0.106 & 1.209 & +0.113 \\
FDK-denoiser + FT & 0.823 & +0.106 & 0.757 & $-$0.025 & 1.372 & +0.086 \\
\bottomrule
\end{tabular}%
}
\end{table*}

\begin{table*}[tbp]
\centering
\caption{Per-axis SSIM: axial value and axial-vs-off-axis gap ($\text{SSIM}_{\text{axial}} - \tfrac12(\text{SSIM}_{\text{coronal}}+\text{SSIM}_{\text{sagittal}})$), at $256^3$/20 views. A gap near 0 indicates the axial plane, in which our 2D model operates, is scored no differently than the two off-axis planes; a positive gap is the SSIM-side signature of inter-slice inconsistency.}
\label{tab:axis_gap}
\resizebox{\textwidth}{!}{%
\begin{tabular}{l|cc|cc|cc}
\toprule
\multirow{2}{*}{\textbf{Method}} & \multicolumn{2}{c|}{\textbf{Dental}} & \multicolumn{2}{c|}{\textbf{Spine}} & \multicolumn{2}{c}{\textbf{Walnut}} \\
\cmidrule(lr){2-3} \cmidrule(lr){4-5} \cmidrule(lr){6-7}
& $\text{SSIM}_{\text{axial}}$ & gap & $\text{SSIM}_{\text{axial}}$ & gap & $\text{SSIM}_{\text{axial}}$ & gap \\
\midrule
FDK & 0.4145 & $-$0.0234 & 0.3594 & $-$0.0323 & 0.1976 & +0.0052 \\
GD$_{zero}$ & 0.7748 & $-$0.0158 & 0.8693 & +0.0020 & 0.5847 & $-$0.0083 \\
GD$_{FDK}$ & 0.7339 & $-$0.0172 & 0.8243 & +0.0054 & 0.5775 & $-$0.0069 \\
\midrule
DPA & 0.8580 & \textbf{+0.0406} & 0.9082 & \textbf{+0.0237} & 0.6499 & +0.0026 \\
CDPA & 0.8942 & +0.0063 & 0.9434 & +0.0059 & 0.7987 & +0.0064 \\
FDK-denoiser & 0.9008 & $-$0.0001 & 0.9510 & +0.0016 & 0.8861 & +0.0000 \\
\midrule
$\mu$(DPA) & 0.9093 & +0.0044 & 0.9451 & +0.0047 & 0.6669 & +0.0009 \\
$\mu$(CDPA) & 0.9202 & $-$0.0018 & 0.9527 & +0.0012 & 0.8135 & +0.0009 \\
FDK-denoiser + FT & 0.8841 & +0.0027 & 0.9434 & $-$0.0001 & 0.7409 & +0.0011 \\
\bottomrule
\end{tabular}%
}
\end{table*}

\subsection{Network Architectures}
We utilize the networks and the optimizer from the diffusers library from Hugging Face~\url{https://huggingface.co/docs/diffusers/en/index}. The diffusion model is based on a U-Net architecture with cross-attention layers, similar to the one used in Stable Diffusion. The network takes as input a noisy image and the time step, and outputs the predicted noise. For the conditional diffusion model (CDPA), we concatenate the FDK reconstruction to the noisy image along the channel dimension, allowing the network to leverage the information from the FDK reconstruction during denoising. The model has cross-attention layers that allow us to attend to the diffusion step, the slice index, and in the case of high-resolution walnut experiments, the number of views.

For the FDK-denoising U-Net model, we use the same U-Net architecture as in the diffusion model, but without the time conditioning. The network takes as input the FDK reconstruction and outputs the denoised image. We also concatenate the slice index to the input along the channel dimension to provide spatial awareness to the network, and use cross-attention on the slice index for conditioning.

We use simple data augmentations during training, including random flips and rotations, to improve the generalization of the models.

We train with the AdamW optimizer with a learning rate of $10^{-4}$ for all models and 1000 diffusion steps for training and 50 for inference. For the guidance, we used 5 epochs with a batch size of $\min\{30, \text{num views}\}$ and a guidance learning rate of $5\times 10^{-4}$.
We used an A100 to train our models, and the training time for the diffusion models was around 24 hours, while for the FDK-denoising U-Net model it was around 6 hours. The inference time is reported in Figure~\ref{fig:ablation_unet} using the same A100 GPU.

\end{document}